\documentclass[twocolumn]{aastex63}

\usepackage{longtable} 
\usepackage{graphicx}
\usepackage{subfigure}
\usepackage{color}

\definecolor{db}{rgb}{0.0, 0.3, 0.8}
\definecolor{dg}{rgb}{0.0, 0.5, 0.0}
\definecolor{dr}{rgb}{0.5, 0.0, 0.0}

\usepackage{lineno}

\def \fexii		{Fe\,{\sc xii}}

\def \arcsec 	{\hbox{$^{\prime\prime}$}}

\received{}
\revised{}
\accepted{}

\submitjournal{ApJ}

\shorttitle{IRIS and EIS \fexii\ observations}
\shortauthors{Del Zanna, G. et al.}

\newcommand{\beq}{\begin{equation}}
\newcommand{\eeq}{\end{equation}}

\begin{document}

\title{Diagnostics of non-Maxwellian electron distributions in solar active regions from \ion{Fe}{12} lines observed by Hinode/EIS and IRIS}

 \correspondingauthor{G. Del Zanna}
\email{gd232@cam.ac.uk}

\author[0000-0002-4125-0204]{G. Del Zanna}
\affiliation{DAMTP, Center for Mathematical Sciences, University of Cambridge, Wilberforce Road, Cambridge, CB3 0WA, UK}

\author[0000-0002-4980-7126]{V. Polito}
\affiliation{Bay Area Environmental Research Institute, NASA Research Park,  Moffett Field, CA 94035, USA}
\affiliation{Lockheed Martin Solar and Astrophysics Laboratory, Building 252, 3251 Hanover Street, Palo Alto, CA 94304, USA}

\author[0000-0003-1308-7427]{J. Dud\'ik}
\affiliation{Astronomical Institute, Academy of Sciences of the Czech Republic, 25165 Ond\v{r}ejov, Czech Republic}

  \author[0000-0002-0405-0668]{P. Testa}
\affiliation{Harvard-Smithsonian Center for Astrophysics,
60 Garden St, Cambridge, MA 02193, USA}

\author[0000-0002-6418-7914]{H.E. Mason}
\affiliation{DAMTP, Center for Mathematical Sciences, University of Cambridge, Wilberforce Road, Cambridge, CB3 0WA, UK}

\author[0000-0003-2629-6201]{E. Dzif{\v{c}}{\'a}kov{\'a}}
\affiliation{Astronomical Institute, Academy of Sciences of the Czech Republic, 25165 Ond\v{r}ejov, Czech Republic}


\begin{abstract}
We present joint Hinode/EIS and IRIS observations of \ion{Fe}{12} lines in active regions, both on-disk and off-limb. We use an improved calibration for the EIS data, and find that the 192.4\,\AA\,/\,1349\,\AA\ observed ratio is consistent with the values predicted by CHIANTI and the coronal approximation in  quiescent areas, but not in all active region observations, where the ratio is often lower than expected by up to a factor of about two. 
We investigate a number of physical mechanisms that could affect this ratio, such as opacity and absorption from cooler material. We find significant opacity in the EIS \ion{Fe}{12} 193 and 195\,\AA\ lines, but not in the 192.4\,\AA\ line, in agreement with previous findings. As we cannot rule out possible EUV absorption by H, He and \ion{He}{2} in the on-disk observations, we focus on an off-limb observation where such absorption is minimal. After considering these, as well as possible non-equilibrium effects, we suggest that the most likely explanation for the observed low \ion{Fe}{12} 192.4\,\AA\,/\,1349\,\AA\ ratio is the presence of non-Maxwellian electron distributions in the active regions. This is in agreement with previous findings based on EIS and IRIS observations independently.
\end{abstract}

 

\section{Introduction}
\label{introduction}

Spectral lines from \ion{Fe}{12} provide a wide range of plasma  diagnostics for the solar corona, as this ion  produces strong lines and is highly abundant. The strongest transitions are in the extreme ultraviolet (EUV), and have been routinely observed by the Hinode Extreme Ultraviolet Imaging Spectrometer (EIS) \citep{culhane_etal:2007}.
Among them, the most intense lines are three decays to the ground state, from $^4$P states of \ion{Fe}{12} , at 192.4, 193.5, and 195.1\,\AA. These \ion{Fe}{12} EIS lines have been widely used for a range of diagnostic applications, especially in active regions.

\begin{figure}
\centering
\includegraphics[angle=90,width=1.\hsize ]{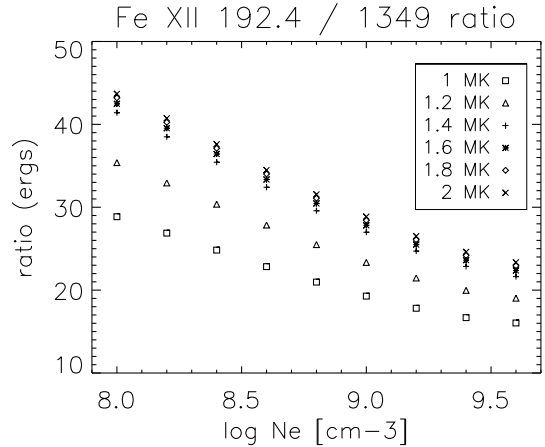}
\caption{Theoretical intensity ratio (ergs) between the EUV 192.4\,\AA\ EIS line and the UV 1349.4\,\AA\ IRIS forbidden line, calculated with CHIANTI v.10 and a range of electron densities and temperatures.}
\label{fig:fe_12_192_1349_tratios}
\end{figure}

\ion{Fe}{12} also produces several weaker forbidden lines in the UV from transitions within its ground configuration.  These include the 1242\,\AA\, line which has been observed by e.g., SoHO SUMER \citep{Wilhelm_etal:1997}, and the 1349\,\AA\ line, observed by the Interface Region Imaging Spectrograph (IRIS) \citep{depontieu_etal:2014}.
Because of the difference in the excitation energies between the  ground configuration states and the upper levels emitting the EUV lines, the ratios of the UV forbidden lines to any of the 192.4, 193.5 and 195.1\,\AA\ lines observed by EIS provide a direct and important diagnostic of the electron temperature, largely independent of any assumption of ionisation equilibrium, although the ratios also have a density dependence as shown in Fig.~\ref{fig:fe_12_192_1349_tratios}.
For the same reason, these ratios are also excellent, unexplored diagnostics for the presence of non-Maxwellian electron distributions (NMED), see e.g., \cite{dudik_etal:2014_fe}. In this case, independent measurements of the electron temperature are necessary. 

We have recently obtained strong evidence that NMED effects are present in active regions \citep{lorincik_etal:2020,dudik_etal:2017}, especially in the active region coronal loops and also the so-called moss, a thin layer at the footpoints of the 3~MK loops \citep{fletcher_depontieu:1999,testa_etal:2013}, where \ion{Fe}{12} emission is brightest \citep[see, e.g.,][]{tripathi_etal:2008,testa_etal:2016}.

The ratios of the \ion{Fe}{12} UV forbidden lines vs.\ the EUV lines is also sensitive to any EUV absorption due to the presence of cool material such as filaments and spicular material, which could significantly affect many diagnostic applications of EUV lines.
Most of the EUV absorption is due to photoionisation of the ground state of neutral hydrogen, with a threshold at 912\,\AA, but significant absorption can also be due to photoionisation of the ground states of neutral helium (threshold at 504\,\AA) and ionised helium (threshold at 228\,\AA).
Such absorption is widespread in the solar corona, and is easily visible in active regions filaments. However, any absorption due to low-lying emission such as spicules is more difficult to measure, as it is inter-mingled with the moss emission. \cite{depontieu_etal:2009_sumer_eis} carried out a comparison between the \ion{Fe}{12} forbidden line observed by SoHO SUMER at 1242\,\AA\ and the 195.1\,\AA\ line observed by Hinode/EIS in an active region. They found that the  195.1\,\AA\,/\,1242\,\AA\  ratio in moss regions was a factor of about 1.5 lower than expected and concluded that a likely explanation for the discrepancy was absorption in the EUV due to cool plasma. They used an early version of CHIANTI, the one which was available at that time. Since then, a large-scale scattering calculation for \ion{Fe}{12} \citep{delzanna_etal:12_fe_12}  significantly changed (by 30--50\%) the populations of the ground configuration states. The new calculations consequently increased significantly the intensities of the forbidden lines. 
The improved  \ion{Fe}{12} atomic data were made available in version 8 of CHIANTI, and are also those in the current CHIANTI v.10 \citep{delzanna_etal:2021_chianti_v10}.
With the improved atomic data, the  195.1/1242\,\AA\ ratio decreases by about a factor of 1.5, bringing them in better agreement with the ratios observed by   \cite{depontieu_etal:2009_sumer_eis} in the moss regions, although not with the loop regions.

%

As IRIS is capable of measuring the  \ion{Fe}{12} 1349\,\AA\ line with a faster cadence than that of SUMER for the 1242\,\AA\ line (about one hour), we devised a Hinode campaign (HOP 246) of simultaneous EIS/IRIS active region observations.
The campaign started on 2014 February 14 and was mostly run in March 2014 on a few active regions, in particular following the disk passage of AR 12014 during the second half of the month.
In spite of relatively long IRIS exposures (30s), the signal for the \ion{Fe}{12} 1349\,\AA\ line, a weak forbidden transition, was consistently low, except for a few observations when the active region was near the limb.
An analysis of two of those observations was presented by
\cite{testa_etal:2016}. Their results focused on Doppler flows and widths, but also indicated a significant discrepancy (up to nearly a factor of two) between the observed and predicted 195.1\,\AA\,/\,1349.4\,\AA~ratios, with the observed ones being systematically higher. The discrepancy increased with the new atomic data in CHIANTI version 8 \citep{delzanna_chianti_v8}, relative to version 7 and seemed to indicate a problem with the atomic data. This was surprising since the benchmarking of \ion{Fe}{12} with observations generally showed good agreement \citep[see a summary in][]{delzanna_mason:2018}. After further investigation, this discrepancy was found, for the most part, to be explained by the errant inclusion of an obsolete keyword in eis\_prep, and the adopted EIS calibration, which is different from the updated version used here. 

To test the \ion{Fe}{12} 192.4\,\AA\,/\,1349.4\,\AA\, diagnostics, we analysed the HOP 246 observations, but also searched the entire IRIS and EIS databases for any other suitable observations where the \ion{Fe}{12} lines were observed by both instruments. We analysed several of these datasets and in the process identified a series of problems associated with the EIS \ion{Fe}{12} observations, as discussed below. 

Section~2 outlines the data analysis and describes some of the main issues we encountered which affected the selection of the observations.
Section~3 describes the observations analysed here, while Section~4 summarizes our conclusions.
An Appendix provides supplementary information.

\section{Data analysis and selection}

\subsection{EIS}
The EIS data were processed with custom-written software \citep[see, e.g.,][]{delzanna_etal:2011_aia}. 
EIS saturation is at 16000~DN, however we found indications of some non-linear effects for lower values approaching this threshold \citep{delzanna_etal:2019_ntw}.
The strongest EIS 195.1\,\AA\ line was sometimes saturated (or close to saturation) in the AR moss regions. For this reason (and for other reasons discussed below) observations of the weaker 192.4\,\AA\ line were used instead.

An analysis  of a large number of EIS observations of different features, on-disk, off-limb, with different exposures, slit combinations, summarised in \cite{delzanna_etal:2019_ntw}, revealed several anomalies in the 192.4, 193.5, and 195.1\,\AA\ lines.  The main ones affect the instrumental widths of the 193.5, and 195.1\,\AA\ lines and their  reduced intensities  (compared to the weaker 192.4\,\AA\ line), in all active region and many off-limb observations. 
The only explanation found for the anomalous ratios and widths of these lines was the ubiquitous presence of opacity effects.
In fact, these three lines are decays to the ground state, so the ratios of these lines are insensitive to density and temperature variations. Their theoretical ratios show agreement with well-calibrated observations of the quiet Sun within 1\% \citep{storey_etal:2005,delzanna_mason:05_fe_12}.
Opacity effects were found to decrease the intensity of the stronger 195.1\,\AA\  line by about 40\%.
Note that in active region observations the relative intensity of the 195.1\,\AA\ line should actually increase (compared to the quiet Sun) due to the presence of a weak \ion{Fe}{12} density-sensitive transition \citep{delzanna_mason:05_fe_12}.

To diagnose the presence of NMED, the temperature needs to be estimated independently.  
The \ion{Fe}{11} lines, identified by \cite{delzanna:10_fe_11} and used in \cite{lorincik_etal:2020} offer such a diagnostic, but are generally not telemetered, and in one of the  observations discussed here are very weak, so we had to resort to standard emission measure analyses.

To measure electron densities and for a meaningful comparison with IRIS, we need to convert the EIS DN values to physical units using a radiometric calibration.
\cite{delzanna:13_eis_calib} presented a significant revision of the ground calibration, with an additional time-dependent decrease in the sensitivity of the long-wavelength channel. That calibration substantially affected the ratios of the  192.4, 193.5, and 195.1\,\AA\ lines, which in quiet Sun on-disk observations were forced to agree with theory and previous observations.
As further wavelength-dependent corrections as a function of time were clearly needed, and the calibration only considered data until 2012, a long-term program was started by GDZ and H.P.Warren, both of the EIS team, to provide an improved radiometric calibration. Here we adopt these new calibration results, as discussed in the Appendix.

 \subsection{IRIS}

The IRIS and EIS observations are generally carried out simultaneously, but are not necessarily co-spatial. In fact, the EIS slit is moved from west to east to `raster' a solar region, while the IRIS slit is moved in the opposite direction \citep[see e.g.,][]{testa_etal:2016}.
Several IRIS observations were carried out with a roll angle of 90 degrees, so that some co-spatial and co-temporal EIS/IRIS observations were guaranteed.  In some instances, several EIS/IRIS rasters were repeated, so it was possible to check the solar variability. 

In addition to the available IRIS and EIS datasets, we also analysed context images using images from the Atmospheric Imaging Assembly \citep[AIA;][]{lemen_etal:2012} telescope on board the Solar Dynamic Observatory \citep[SDO;][]{pesnell_etal:2012} in the 193\,\AA\ broad band filter,  to select observations with small solar variability (we note that the AIA 193\,\AA\  band is typically dominated by the three \ion{Fe}{12} 192.4, 193.5, and 195.1\,\AA\ lines in the moss regions; e.g., \citealt{MartinezSykora2011}).

IRIS level 2 data were used. The data were spatially binned as described below, to improve the signal. The \ion{Fe}{12} line was fitted with a Gaussian in each pixel and the conversion to physical units was performed afterwards. 
The radiometric calibration of the IRIS instrument is discussed in \cite{wuelser_etal:2018}.
The uncertainty in the IRIS calibration for the data analysed here is of the order of 20--30\%

We also note that \cite{testa_etal:2016} showed that in cases with low signal-to-noise (especially when the peak DN in the line is less than 10), the intensity of the line is likely to be under-estimated by up to $\sim 15$ percent. 
For the comparisons with EIS, we typically only consider the regions where the IRIS line has averaged peak DN above 20. Hence, we have not applied any such corrections to the IRIS intensities.

During the analysis of the on-disk observations, we noticed the presence of an unidentified photospheric (very narrow) line at exactly the rest wavelength of the 1349.4\,\AA\ \ion{Fe}{12} line (see the Appendix). We have estimated that the contribution of this line to on-disk moss regions is however minimal, of the order of 5\% in some locations, by using the  \ion{C}{1} 1354.28\,\AA\ line as a reference.

In addition, we estimate that the theoretical emissivity of the 1349.4\,\AA\ \ion{Fe}{12} line is accurate to within 10\%.  
As mentioned, the new atomic model (CHIANTI, v8) increased the intensities of the forbidden lines by 50 percent or more. A benchmark of the v.8 atomic data against quiet Sun off-limb SoHO SUMER observations 
indicated excellent agreement \citep{delzanna_deluca:2018}. 
The population of the upper state of the 1349.4\,\AA\ line is mainly driven by cascading effects.
Improvements with future atomic calculations cannot be ruled out.
However, it is unlikely that larger calculations would affect the line by more than a few percent.
In the main scattering calculations, all states up to $n=4$ were included by \cite{delzanna_etal:12_fe_12}. Cascading effects from higher states up to the main $n=6$ were included with an approximate (distorted wave) calculation, showing an increase in the forbidden lines by about 3\%. These cascading effects were not included in CHIANTI v8 as the size of the model ion would have been very large, and as a 3\% increase was deemed negligible.

\subsection{On-disk observation - 2014 March 29}
\label{sec:obs}

Following the above-mentioned data selection constraints, we analysed several observations.
A large scatter in the 192.4\,\AA\,/\,1349.4\,\AA~ratios was found, although consistent results among the various measurements were also found. We provide results for one of the on-disk observations, that obtained on 2014 March 29 by IRIS at 23:27-02:14 UT, i.e., the same observing sequence analyzed by \cite{testa_etal:2016}.

\begin{figure*}[!htbp]
\centering
\includegraphics[angle=0,width=1.0\hsize ]{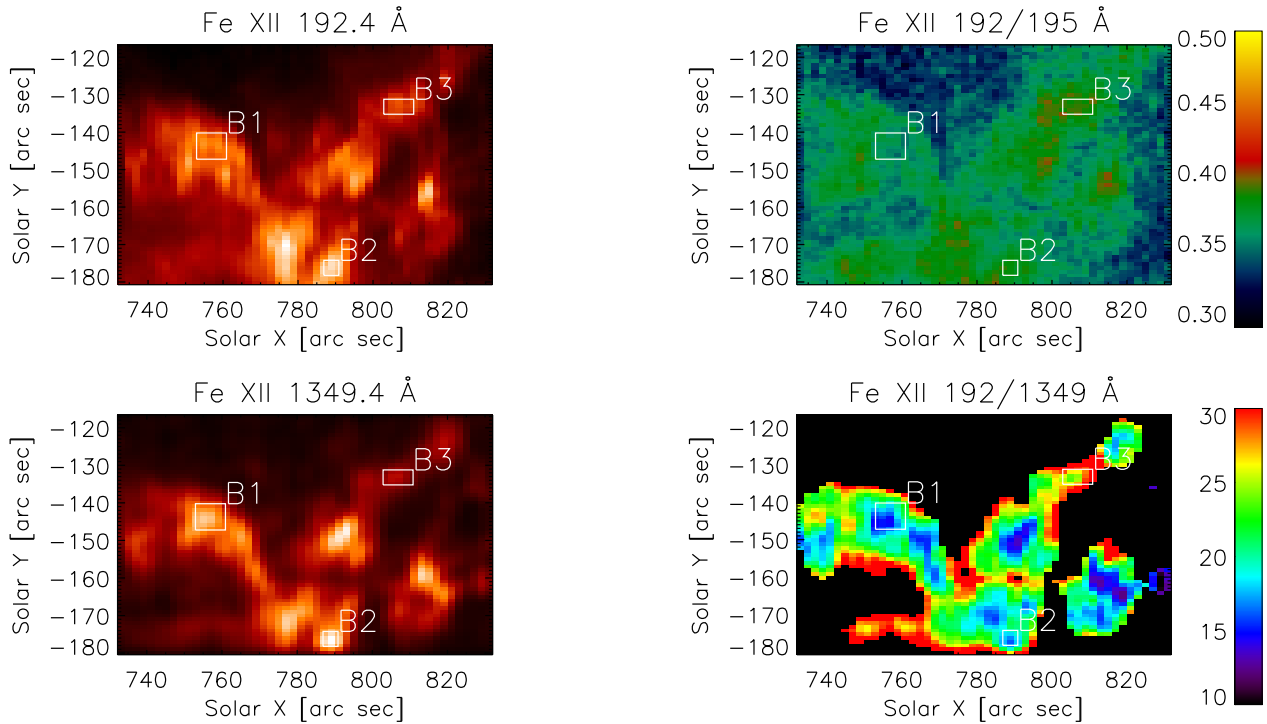}
\caption{Top: intensity image in the \ion{Fe}{12} 192.4\,\AA\ line and ratio with the 195.1\,\AA\ line (ergs) for the 2014-03-29 observation. 
Note that the ratio should have a value of 0.31. Bottom: intensity image of the IRIS 1349.4\,\AA\ line and the 192.4\,\AA\,/\,1349\,\AA\ ratio (ergs).
}
\label{fig:29_mar_2014_images}
\end{figure*}

The EIS raster we focus on here was obtained during 23:24 and 23:50 UT. Note that in \cite{testa_etal:2016} a later raster obtained a couple of hours later (2014 March 30 01:36-02:02) was analyzed.
In the brightest moss regions, the EIS 195.1\,\AA\ line reached 15,000 DN, i.e., was very close to saturation.
Fig.~\ref{fig:29_mar_2014_images} (top) shows an image of the integrated intensity of the \ion{Fe}{12} 192.4\,\AA\ line and its ratio (ergs) with the 195.1\,\AA\ line. The expected ratio is 0.31, which is generally observed on-disk, but not in the brightest moss regions, where the ratio increases to values around 0.4, an indication of some opacity effects.
We have assumed  that the opacity effects in the 192.4\,\AA\ line are negligible (see discussion below), and used this line for the comparison with IRIS.

The IRIS raster was obtained with a -90$^o$ roll angle, 30s exposures and stepping the 0.33\arcsec\ slit by 1\arcsec, with 64 steps.
The IRIS data were rotated and rebinned by a factor of 12 along the slit, to obtain a spatial resolution in the EW direction comparable to the EIS one, as EIS rastered with about 2\arcsec\ steps using the 2\arcsec\ slit.
In the other direction, the IRIS data were first interpolated linearly over a 0.33\arcsec\ grid, and then rebinned by a factor of 3, to achieve a spatial resolution of 1\arcsec, equivalent to the EIS pixel size along the slit. 

The contribution of the unidentified cool line blending the IRIS \ion{Fe}{12} line was estimated by removing 4\% of the \ion{C}{1} line at 1357.13\,\AA.
This resulted in a small correction, of the order of 5 percent in a few places, i.e., not affecting the main results.

As the effective spatial resolution of EIS is about 3--4\arcsec (partly due to the jitter during the long exposures), for a direct pixel-to-pixel comparison, the IRIS data were convolved to obtain an effective spatial resolution close to the EIS one. { 
Such smoothing was not carried out in the analysis by \cite{testa_etal:2016},
which may explain why a broader scatter in the ratios was found in their analysis, compared to what is shown here.}  Finally, the EIS and IRIS images were co-aligned by cross-correlation.
The resulting IRIS image is shown in Fig.~\ref{fig:29_mar_2014_images} (bottom), together with the calibrated ratio of the 192.4\,\AA\,/\,1349.4\,\AA\ lines (in ergs).
It is clear that a pixel-to-pixel comparison has some limitations, as in some places the morphology in the EIS and IRIS lines is not quite the same. That is partly due to the non-simultaneity, partly due to the EIS effective resolution which is very difficult to model. 
However, overall the comparison is satisfactory. 
Figure~\ref{fig:29_mar_2014_images} shows that the 192.4\,\AA\,/\,1349.4\,\AA\ ratio varies significantly, between values close to 30 in some regions to around 15 in the brightest regions.

The 192.4\,\AA\,/\,195.1\,\AA\ ratio (shown in Fig.~\ref{fig:29_mar_2014_images}) is indicative of some opacity effects, which would be significant in the 195.1\,\AA\ line, but relatively small (about 10\%) in the weaker  192.4\,\AA\ line (see a discussion below on opacity issues).

\begin{figure}
\centering
\includegraphics[angle=0,width=.98\hsize ]{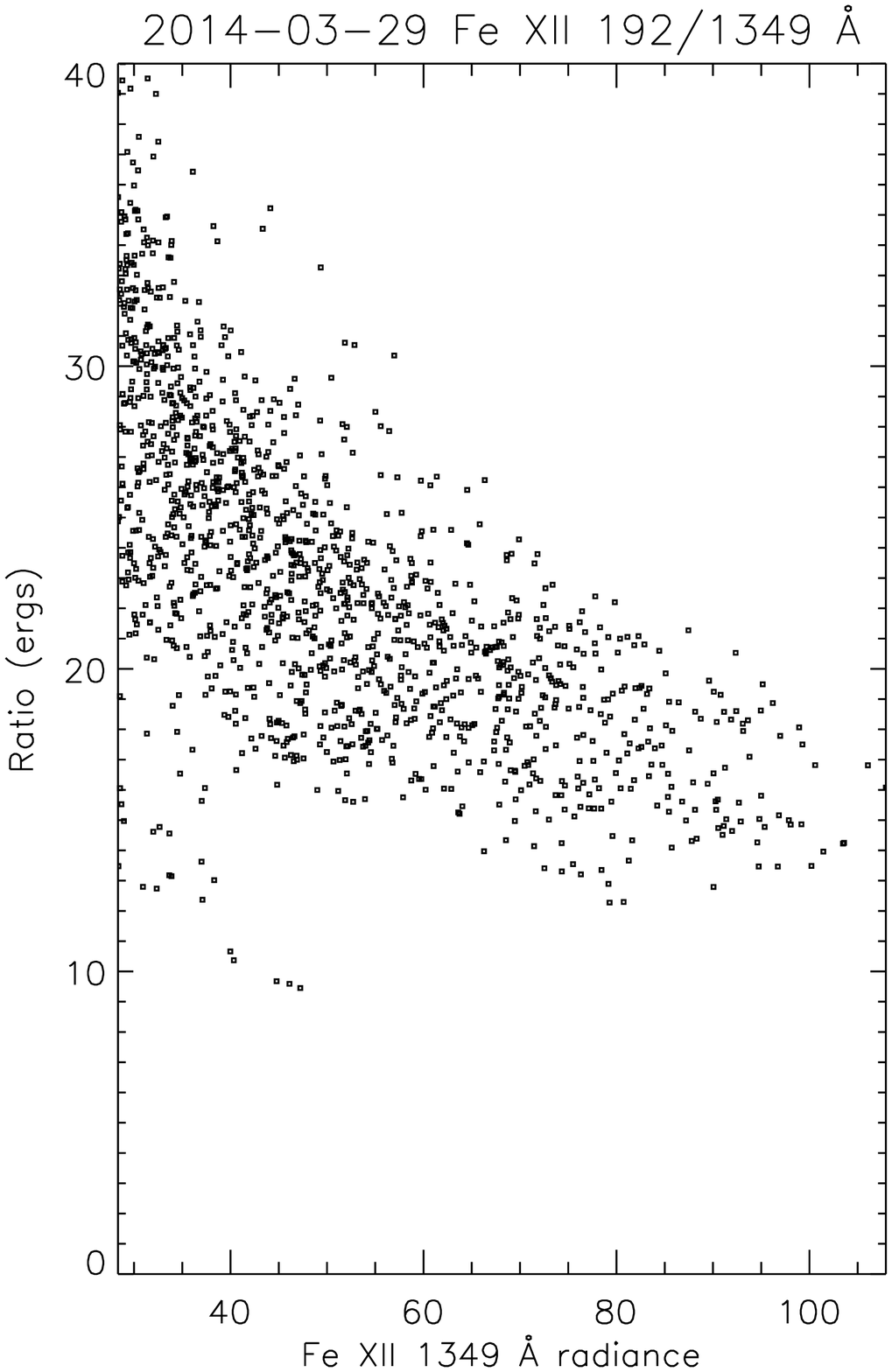}
\caption{The 192.4\,\AA\,/\,1349.4\,\AA\ ratio for the  observation on 2014-03-29,
as a function of the calibrated intensity in the IRIS line.}
\label{fig:29-mar-ratio}
\end{figure}

\begin{figure}
\centering
\includegraphics[angle=270,width=1.\hsize ]{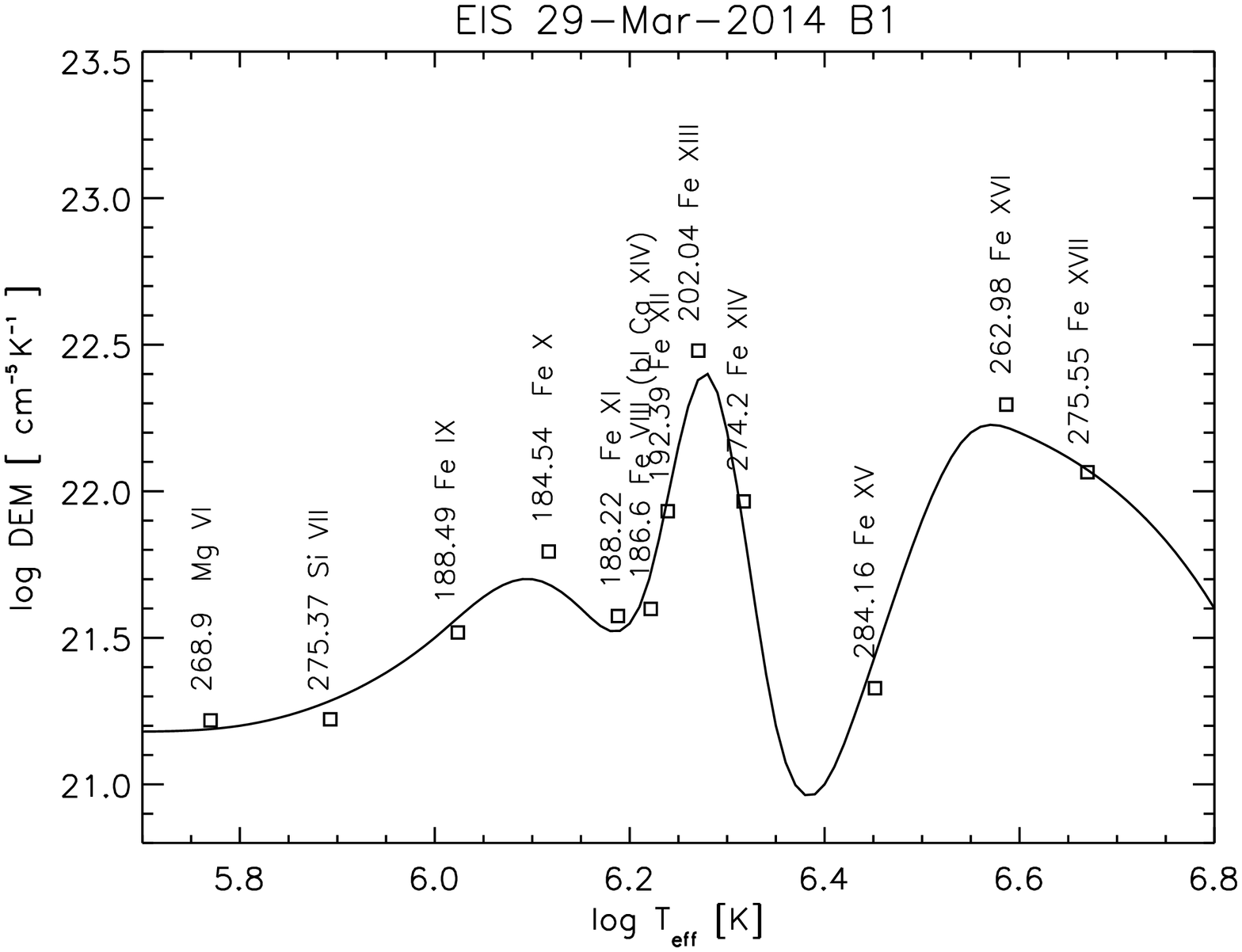}
\includegraphics[angle=270,width=1.\hsize ]{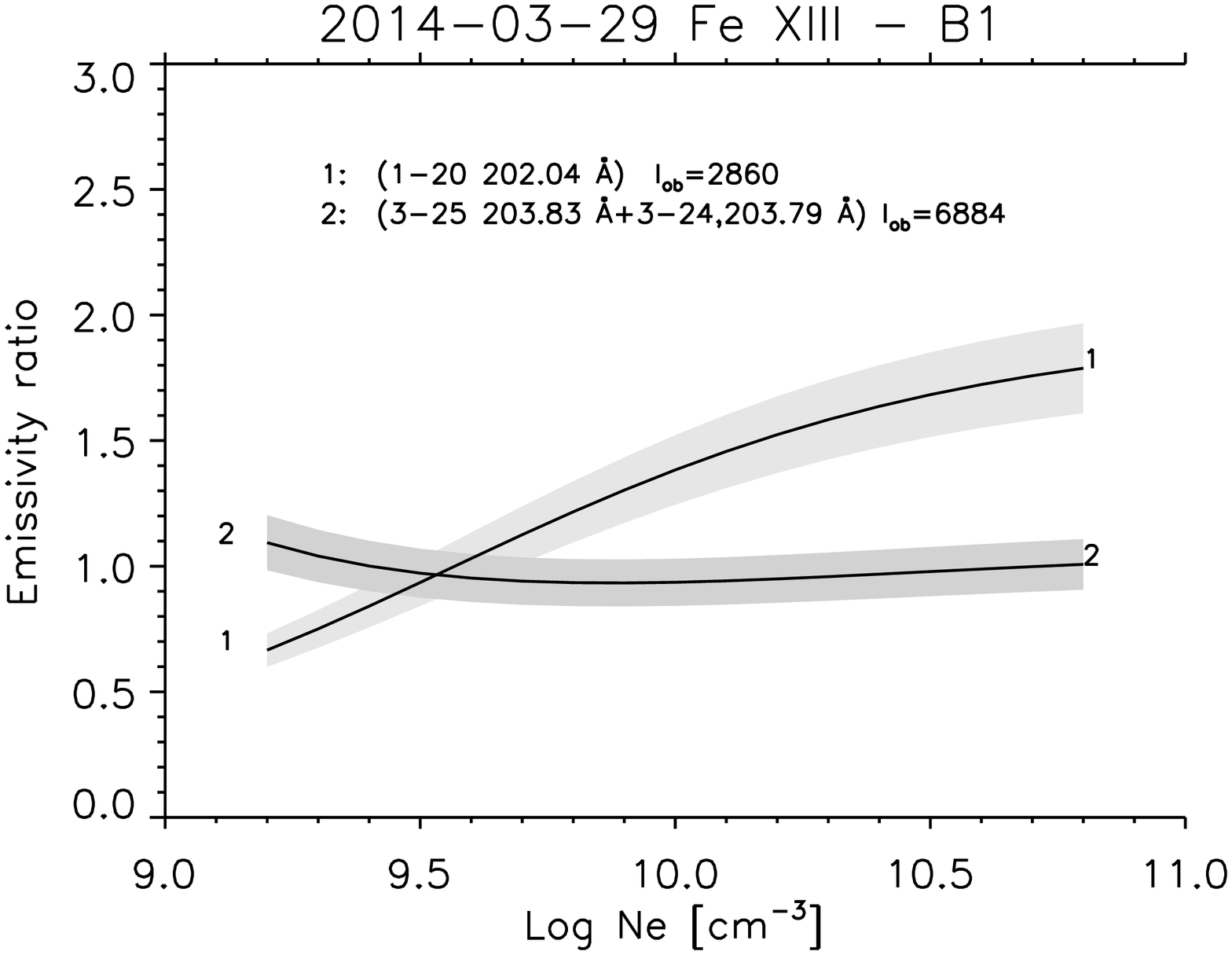}
\caption{Top: DEM for the B1 region, indicated in Fig.~\ref{fig:29_mar_2014_images}. The points are plotted at the effective temperature, and at the
theoretical vs. the observed intensity ratio multiplied by the DEM value. The wavelengths (\AA) and main ion are indicated.
Bottom: emissivity ratio of the main \ion{Fe}{13} lines in the 
B1 region.}
\label{fig:29-mar-em}
\end{figure}

A scatter plot of the 192.4\,\AA\,/\,1349.4\,\AA\ ratio as a function of the calibrated intensity in the IRIS line is shown in Fig.~\ref{fig:29-mar-ratio}. 
It shows a large variation of about a factor of two, with lower values where \ion{Fe}{12} is brightest, in the moss regions.
We selected three moss regions, indicated as B1, B2, and B3 in 
Fig.~\ref{fig:29_mar_2014_images} and measured the averaged density.
The averaged intensities (obtained from the pixel-by-pixel
  measurements) in the lines, their ratios and the averaged densities
  are shown in Table~\ref{tab:29_mar_2014}. The averaged density is about   3 $\times 10^9$ cm$^{-3}$ using the \ion{Fe}{13} lines. 
  The densities from \ion{Fe}{12} are higher, partly because 
  of opacity effects.
We then measured the temperature distribution with both an EM loci and a DEM method, using coronal abundances. For the DEM analysis we used a 
modified version of the CHIANTI v.10 programs, where the DEM is modelled as a spline function and the routine MPFIT is used.  The DEM results for the region B1 are shown in Fig.~\ref{fig:29-mar-em} as an example. 
The temperature distribution is multi-thermal, but the \ion{Fe}{12} and \ion{Fe}{13} lines can also be reasonably modelled with an isothermal plasma around 2 MK. 
In the moss region B1, the averaged ratio is about 19, lower than 25.1, the expected value calculated with the measured density and the DEM.

Note that this is the same AR observed in \cite{testa_etal:2016} (although not observed at the same time; see sec.~\ref{sec:obs}), where we recall the 195.1\,\AA\,/\,1349.4\,\AA\ ratios were found {\it higher} than predicted (up to nearly a factor of two). We tracked down the reason for this large difference, which was mostly due to an obsolete keyword ({\sc correct\_sensitivity}) in the EIS standard {\sc eis\_prep} software, and the different EIS calibration used.

The large variations in the ratio and the very low values (about 15) 
need to be explained. They could be due to strong EUV absorption by 
neutral hydrogen and helium emission 
or other non-equilibrium effects discussed below. 
Cool filamentary material is always present in active regions, but its 
absorption is difficult to quantify, unless it is higher up in the corona
and the underlying emission can reliably be estimated. 
In this observation, and the other ones we have analysed, we did not
find obvious evidence that the lower ratios were due to cool filaments.
However, we cannot rule out the possibility that 
neutral hydrogen and helium is intermixed with the moss emission. 

\begin{table*}[!htb]
  \caption{Intensities I (ergs) and ratios R (ergs)
    in the moss regions observed  on 2014-03-29.
Values in parentheses are DN. The last column shows the densities from the \ion{Fe}{13} 204\,\AA\ / 202\,\AA\ ratio.}
\centering
  \begin{tabular}{lllllllll}
  \hline
Region   & I (192\,\AA) & I (195\,\AA) & I (1349\,\AA) & R (192/195\,\AA) & R (192/1349\,\AA) & Log Ne \\
\hline
B1 & 1480 (11905) & 3970 (45533) & 78 (1100) & 0.37 & 19.0 & 9.5 \\
B2 & 1850 (14881)  & 4630 (53015) & 94 (1330) & 0.40 & 19.7 & 9.45 \\
B3 & 1280 (10274) & 3180 (36400) & 41 (581)  & 0.40 &  31.2 & 9.5 \\
\hline 
\end{tabular}
\label{tab:29_mar_2014}
\end{table*}

We have analysed other on-disk observations of active regions, and 
found similar results to those shown above. 
Aside from other observations of the same AR at the end of March 2014,
we have analysed in detail an observation on 2013 Oct 7
and one on 2013 Nov 30.

\subsection{Off-limb observation of 2013-10-22}

To reduce the possible effects of absorption by cool material, we have searched for off-limb observations with minimal filament material. Unfortunately, only one suitable observation was found. This was obtained on 2013-10-22. 
The EIS study designed by us (cam\_ar\_limb\_lite\_v2) was rich in diagnostic lines and had a good signal, as the exposure was 45\,s.
One problem with this observation was the presence of a storm of high-energy particles so each exposure had to be inspected to remove those particle hits, as standard cosmic ray removal procedures did not work. In spite of this, some anomalous intensities are still present due to residual particle hits/warm pixels in some weaker lines.
EIS rastered during 06:45--8:51 with the 2\arcsec\ slit an off-limb region where a small active region was present. Most of the AR was located well behind the east limb, as we could judge from AIA observations of the following days.

We checked for the presence of cool filaments or spicular material using AIA observations in 304\,\AA, but also 193\,\AA~and 211\,\AA, together with H$\alpha$ observations by the Kanzelh\"{o}he Observatory. 
The co-alignment of AIA with EIS was achieved using a slicing method for the AIA 193\,\AA~data to produce a pseudo-raster corresponding to EIS \ion{Fe}{12} 192.4\,\AA. We find that best co-alignment is found if the AIA is rotated with respect to EIS by about 0.5$^\circ$, as well as shifted by a few arc seconds in both axes. The Kanzelh\"{o}he H$\alpha$ data  traditionally have excellent pointing, which we verified by comparison with AIA 193\,\AA, focusing on filaments off-limb. Thus, the Kanzelh\"{o}he H$\alpha$ data were coaligned with EIS analogously to AIA data.

\begin{figure*}[!htbp]
\centering
\includegraphics[width=6.64cm,viewport= 0 40 331 360,clip]{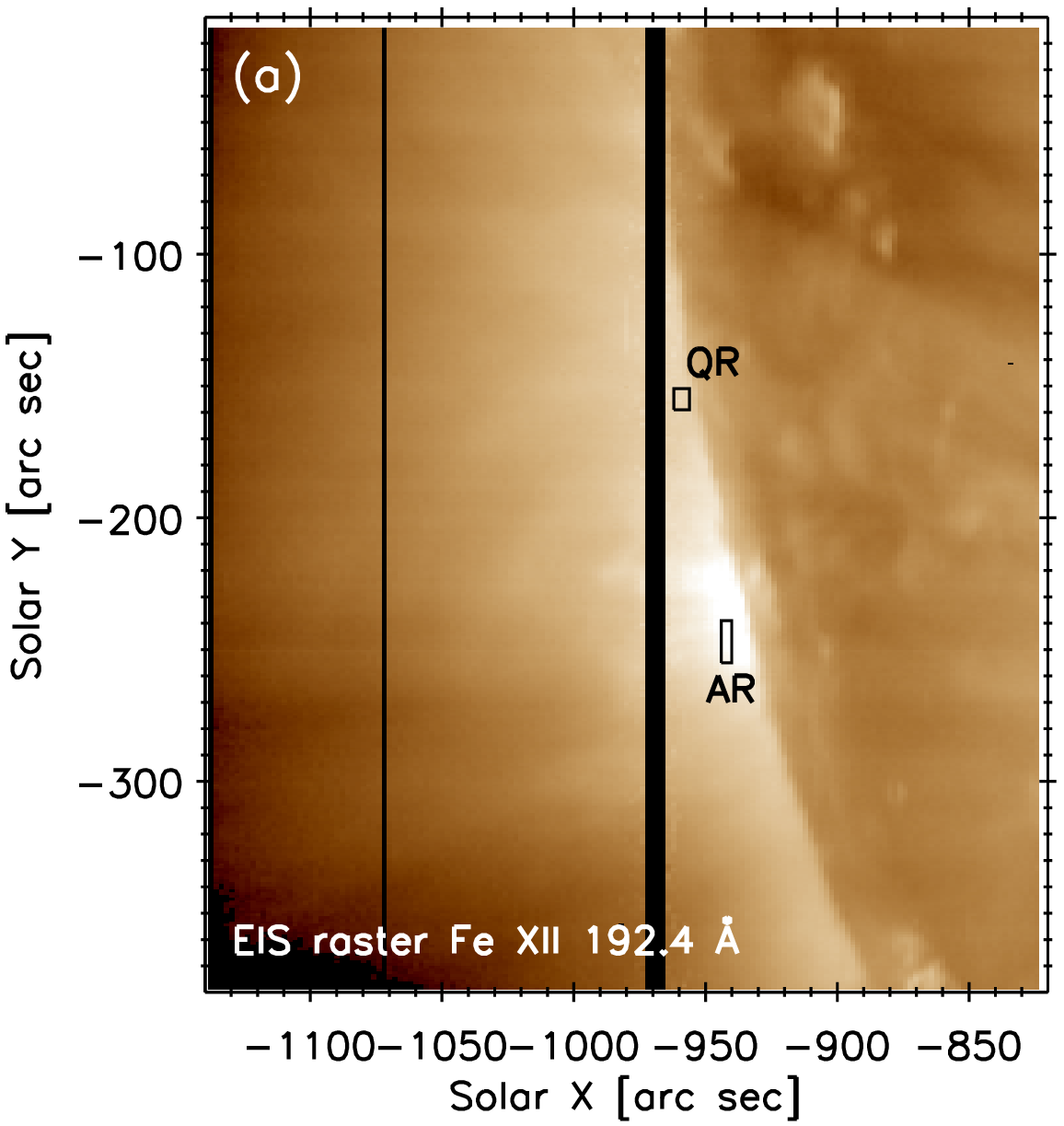}
\includegraphics[width=5.47cm,viewport=58 40 331 360,clip]{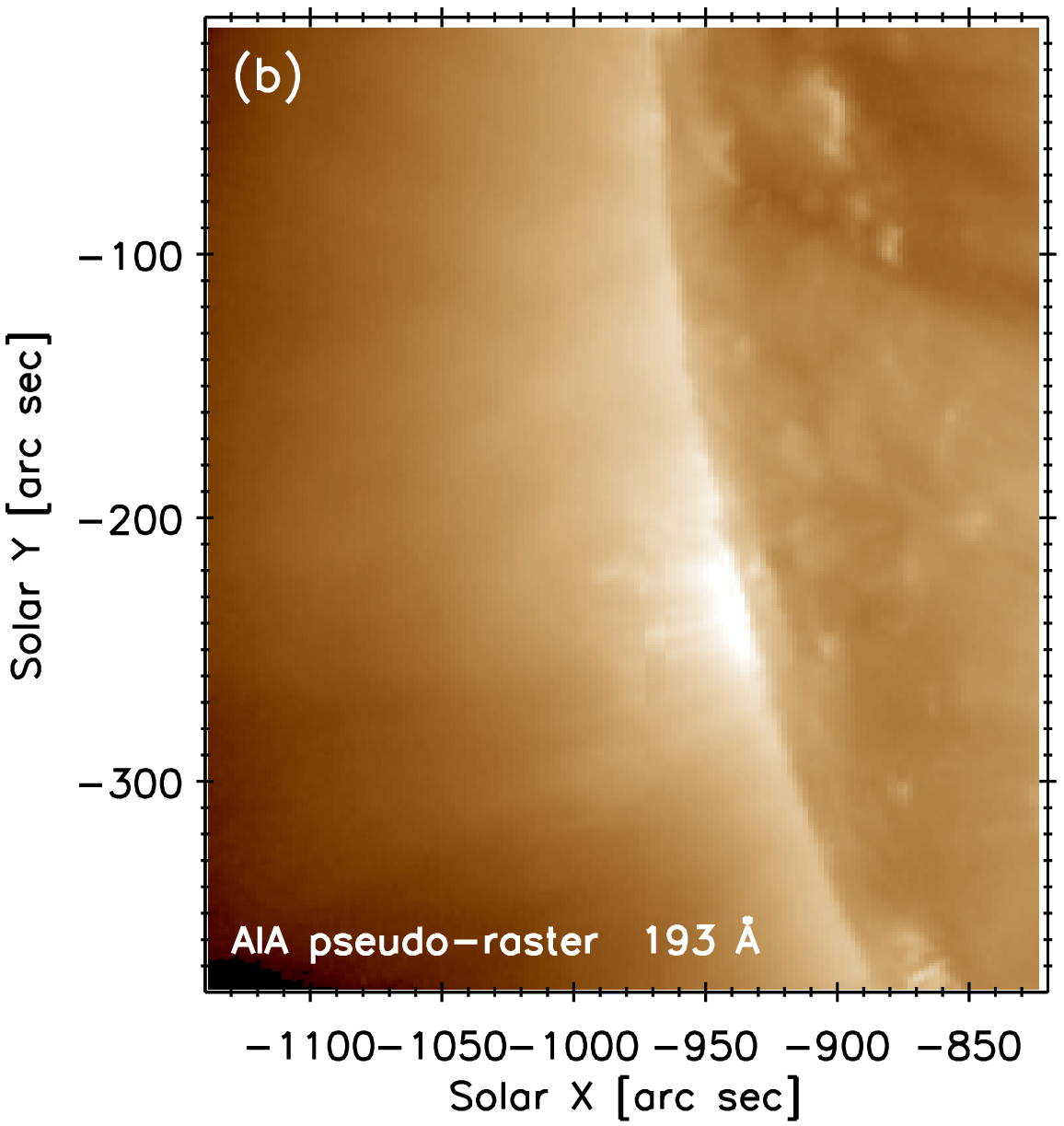}
\includegraphics[width=5.47cm,viewport=58 40 331 360,clip]{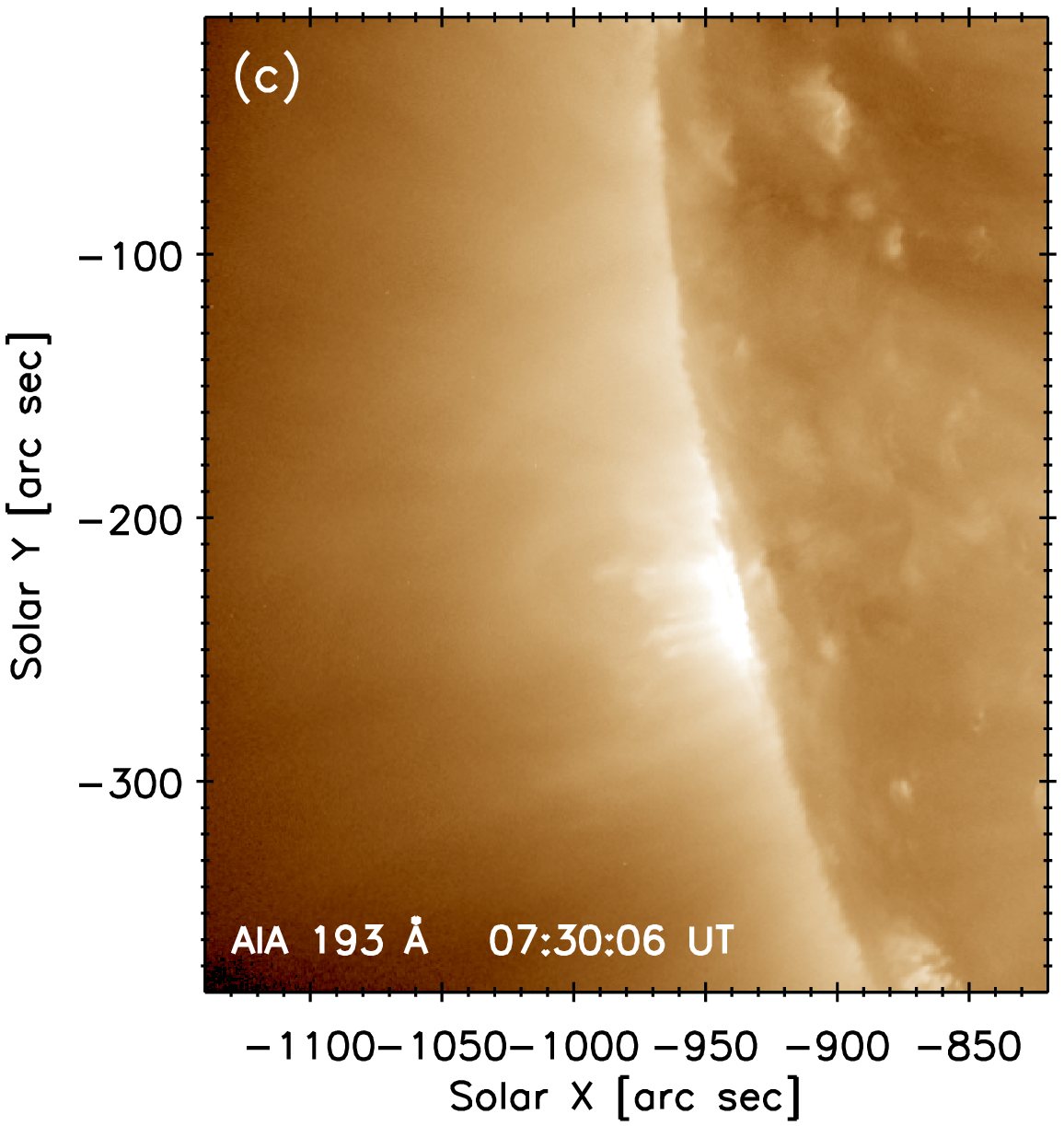}
\includegraphics[width=6.64cm,viewport= 0 0 331 360,clip]{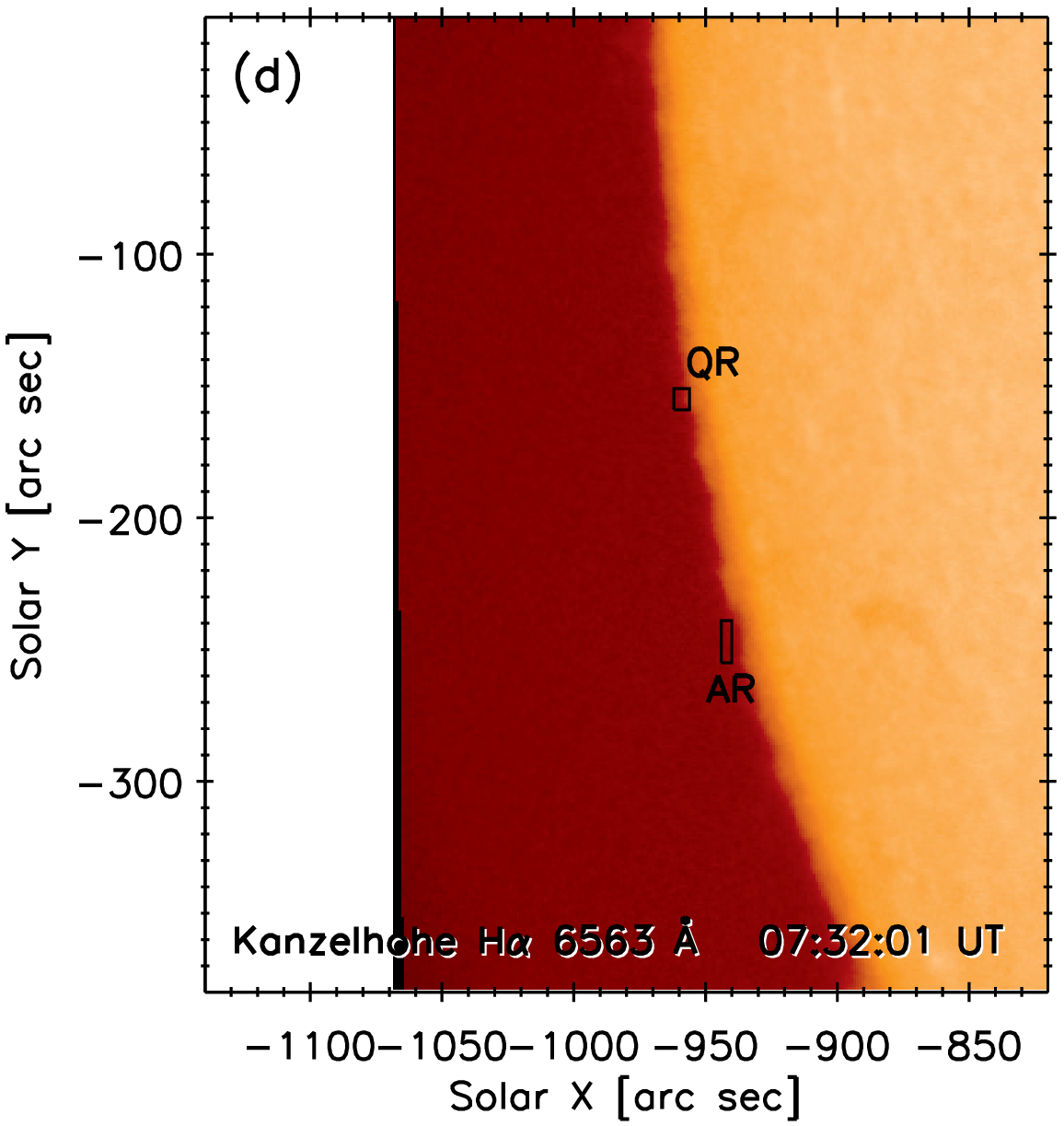}
\includegraphics[width=5.47cm,viewport=58 0 331 360,clip]{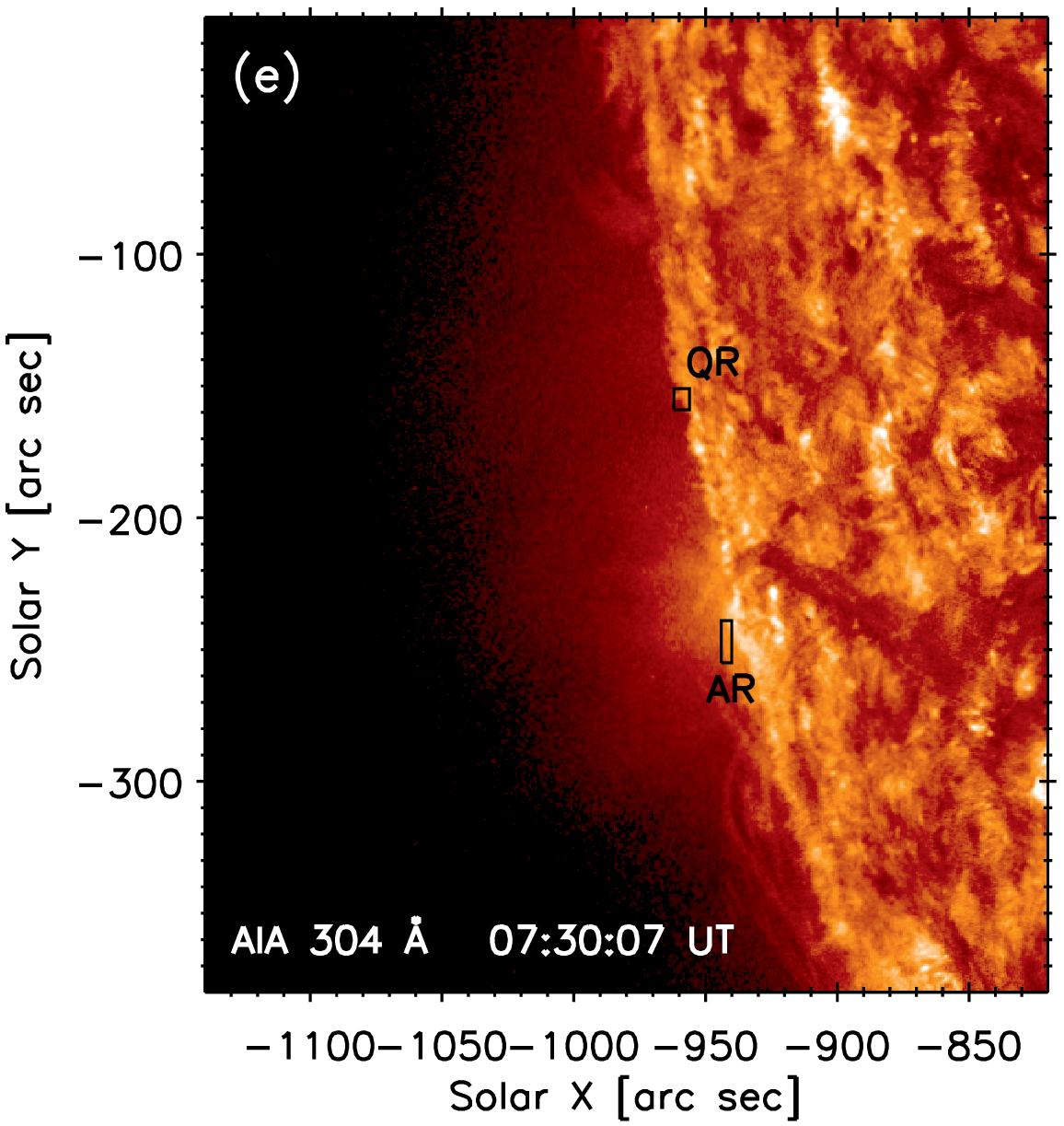}
\includegraphics[width=5.47cm,viewport=58 0 331 360,clip]{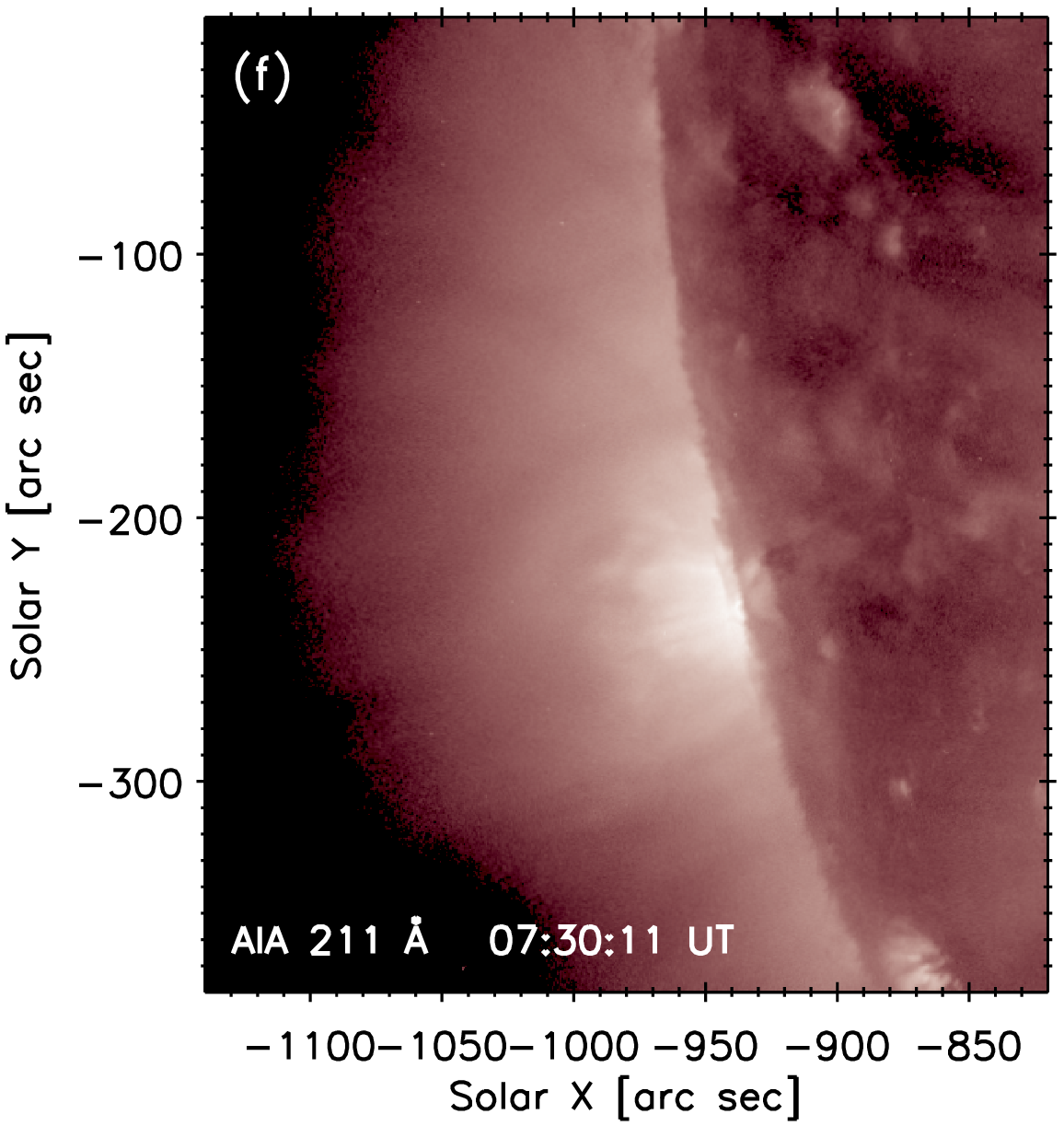}
\caption{Context observations of the 2013-10-22 off-limb active region. The EIS \ion{Fe}{12} 192.4\,\AA~line is shown in the panel (a), while the AIA 193\,\AA~pseudo-raster is shown in panel (b). Panels (c)--(f) show snapshots from AIA and Kanzelhohe H$\alpha$ observations, all coaligned to match the EIS and IRIS observations.}
\label{fig:2013_10_22_aia}
\end{figure*}

The context AIA and H$\alpha$ data are shown in  Figure \ref{fig:2013_10_22_aia} alongside the EIS raster. We have selected two regions for further analysis, which are labelled as 'AR' and 'QR'.

The H$\alpha$ data and the AIA coronal images do not show any 
indications of absorption by cool material off-limb in these regions. 
The main absorption would be due to neutral hydrogen and neutral helium,
with a minor contribution from ionized helium.

The AIA 304\,\AA\ images  show some emission above the limb in the `AR' region, but the amount of ionized helium is difficult to quantify, for multiple reasons, including uncertainties in the chemical abundances, instrument calibration, and coronal contribution to the band.  
We estimate that in the `AR' region the Si XI 303.33\,\AA\  line alone accounts for about a quarter of the AIA count rates (which are about 40 DN\,s$^{-1}$).
In fact, with the DEM distribution we obtained, the intensity of the Si XI line results 5280 (erg s$^{-1}$ cm$^{-2}$ sr$^{-1}$). Using the estimated effective area of the AIA channel (for this observation and normalised to EVE), that is equivalent to an average of 11 DN\,s$^{-1}$ per AIA pixel due to Si XI. We note that the resonance \ion{He}{2} lines at 303.8\,\AA~are formed at higher temperatures and have much larger optical thickness than H$\alpha$, which in turn has similar optical thickness to the H and He continua around 195\,\AA~\citep[e.g.][]{wang:1998,anzer_heinzel:2005}.
Thus, the presence of weak-signal structures in \ion{He}{2}, but not in H$\alpha$ along the LOS is still consistent with negligible absorption of EUV radiation by chromospheric or transition-region material.

IRIS scanned the same region from east to west with the 0.33\arcsec\ slit, 30s exposure times and `sparse rastering', i.e., the slit location was stepped by 1\arcsec.
The interesting area above the limb, where some IRIS signal from \ion{Fe}{12} was present, was observed almost simultaneously by IRIS and EIS. We performed the IRIS and EIS data reduction and calibration in a similar way to that described in the previous section.

\begin{figure}
\centering
\includegraphics[angle=0,width=0.453\hsize,bb= 50 0 260 566,clip]{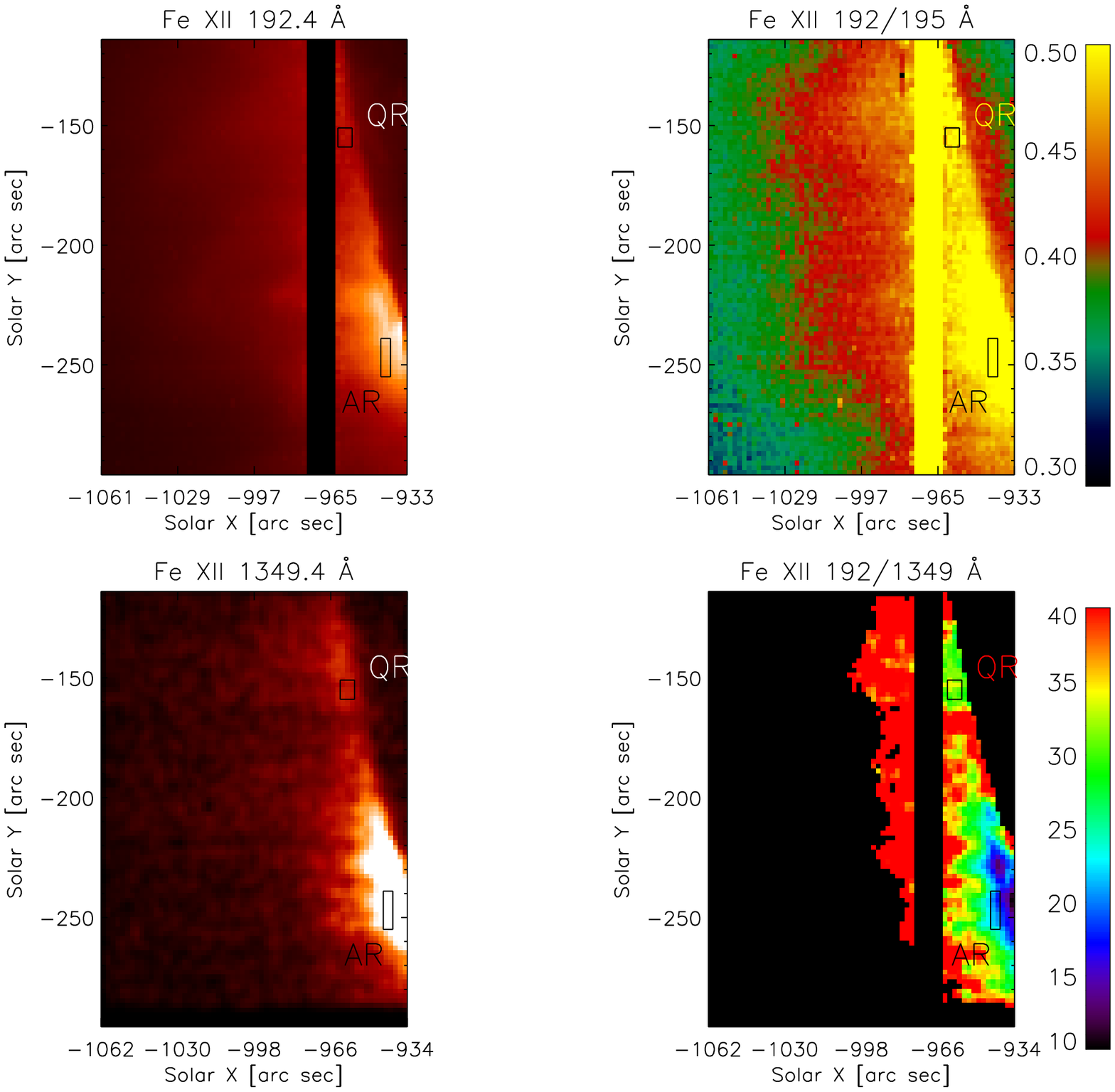}
\includegraphics[angle=0,width=0.527\hsize,bb=375 0 620 566,clip]{22_oct_2013_images.eps}
\caption{Summary of the EIS/IRIS comparison on the off-limb AR observation on 2013-10-22.}
\label{fig:eis_iris_ratio2}
\end{figure}

\begin{figure}
\centering
\includegraphics[angle=0,width=.98\hsize ]{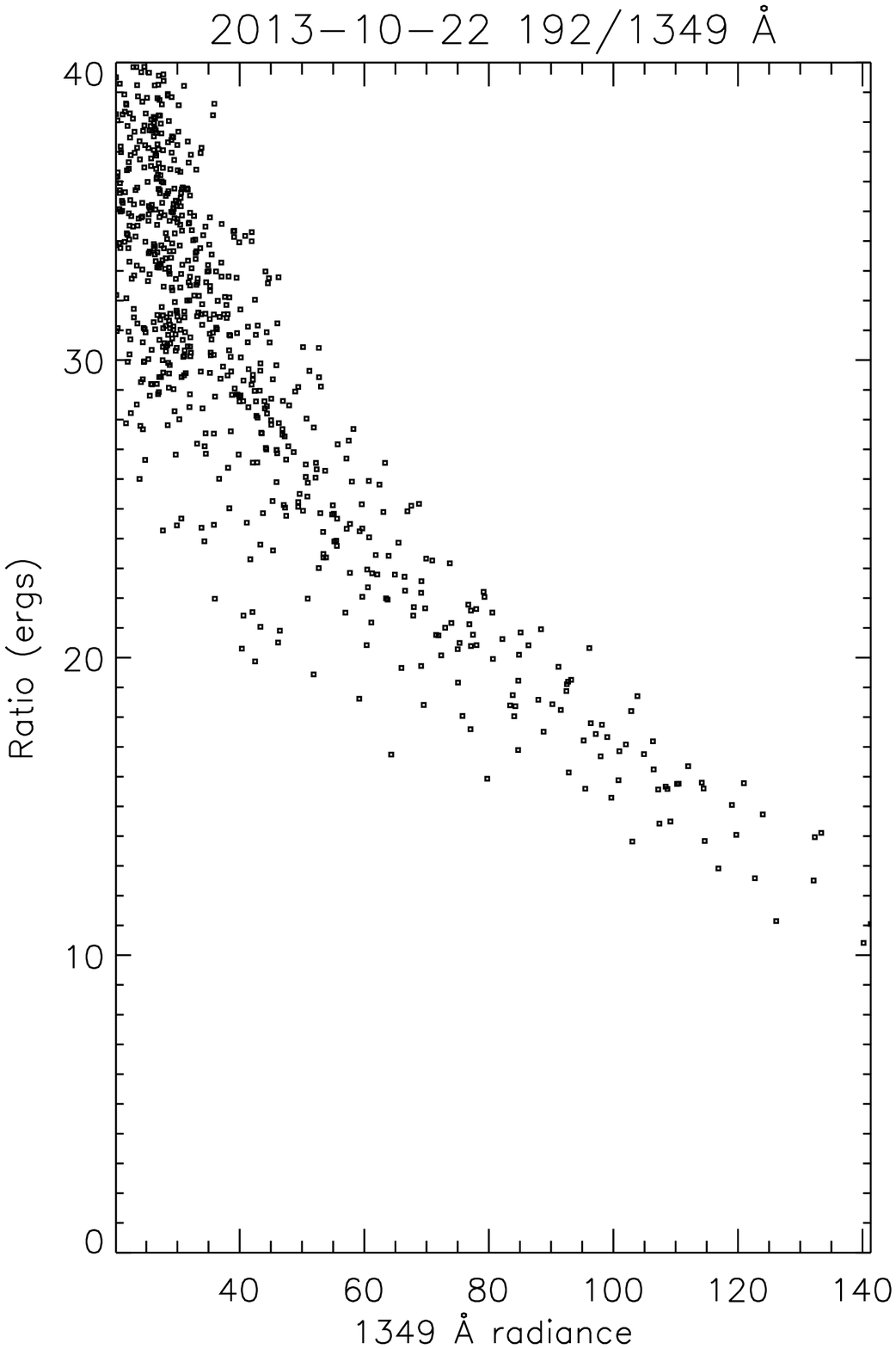}
\caption{Scatter plot for the off-limb AR observation
on 2013-10-22.}
\label{fig:eis_iris_ratio_sp2}
\end{figure}

Figure~\ref{fig:eis_iris_ratio2} shows a summary of the EIS/IRIS comparison. As for the on-disk cases, the 192.4/195.1\,\AA\ is higher in the brightest regions, indicating some opacity effects. The width of the 195.1\,\AA\ line is also larger in the same regions. As in the previous cases, the 192.4\AA\,/\,1349.4\,\AA~ratio varies significantly from values around 30, north of the AR, to values around 15 closest to the core of the AR.
Figure~\ref{fig:eis_iris_ratio_sp2} shows the scatter plot of this ratio.
Averaged intensities and ratios in those regions are shown in Table~\ref{tab:22_oct_2013}. 

\begin{table*}[!htb]
  \caption{Intensities I (ergs) and ratios R (ergs) in the two the off-limb regions observed on 2013-10-22.
  Values in parentheses are intensities in DN (the exposure times for EIS and IRIS were 45 and 30 seconds, respectively). }
\centering
\begin{tabular}{lllllllll}
  \hline
Region   & I (192\,\AA) & I (195\,\AA) & I (1349\,\AA) & R (192/195\,\AA) & R (192/1349\,\AA) & \\
\hline
AR  &  1545 (19412)  & 2920 (51907) &  73 (3343) & 0.53 & 21 & \\
QR  &    880 (11049)  & 1770 (31521) &   28.1 (1288) & 0.50 & 31 & \\
\hline 
\end{tabular}
\label{tab:22_oct_2013}
\end{table*}

\begin{figure}
\centering
\includegraphics[angle=-90,width=.9\hsize ]{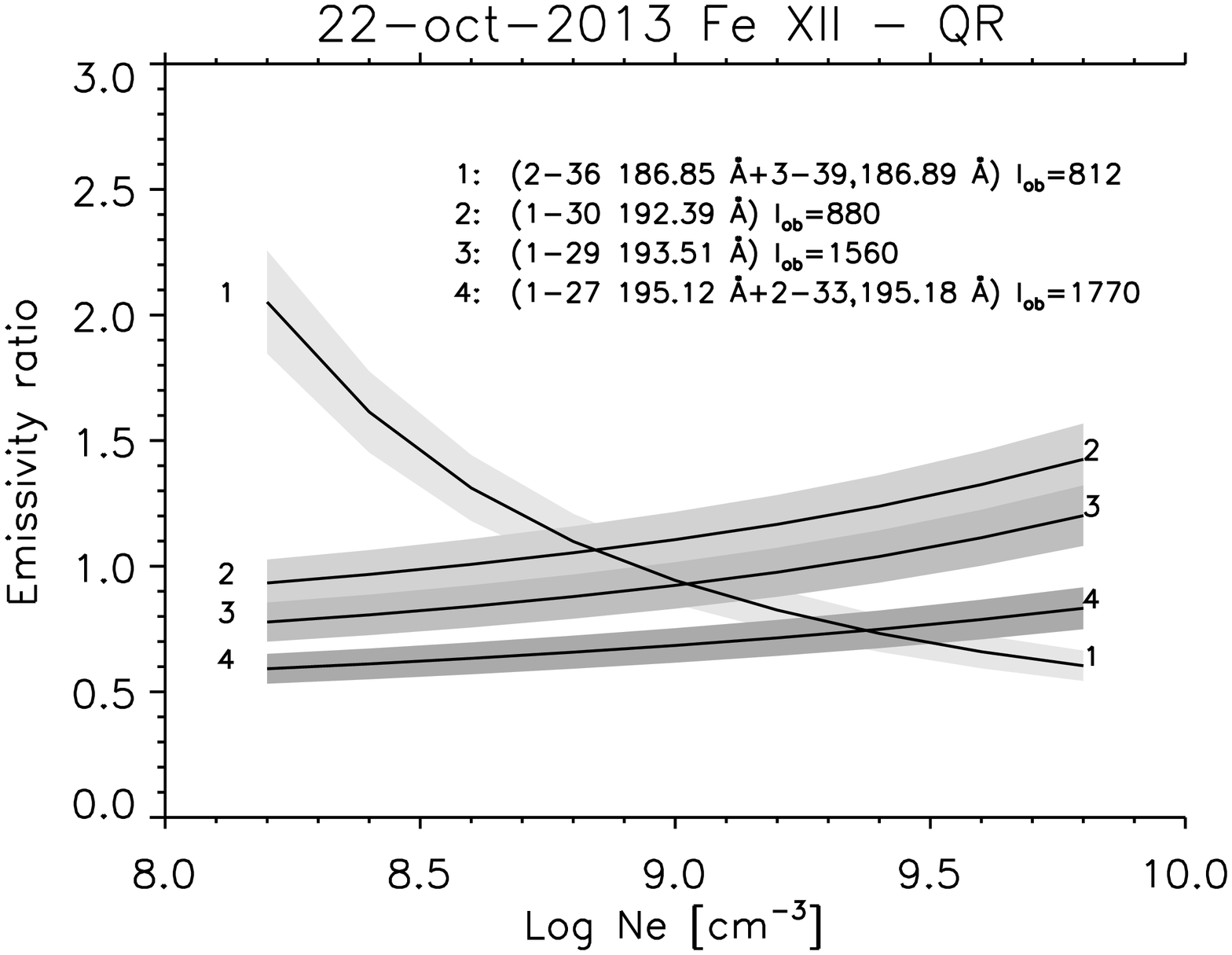}
\includegraphics[angle=-90,width=.9\hsize ]{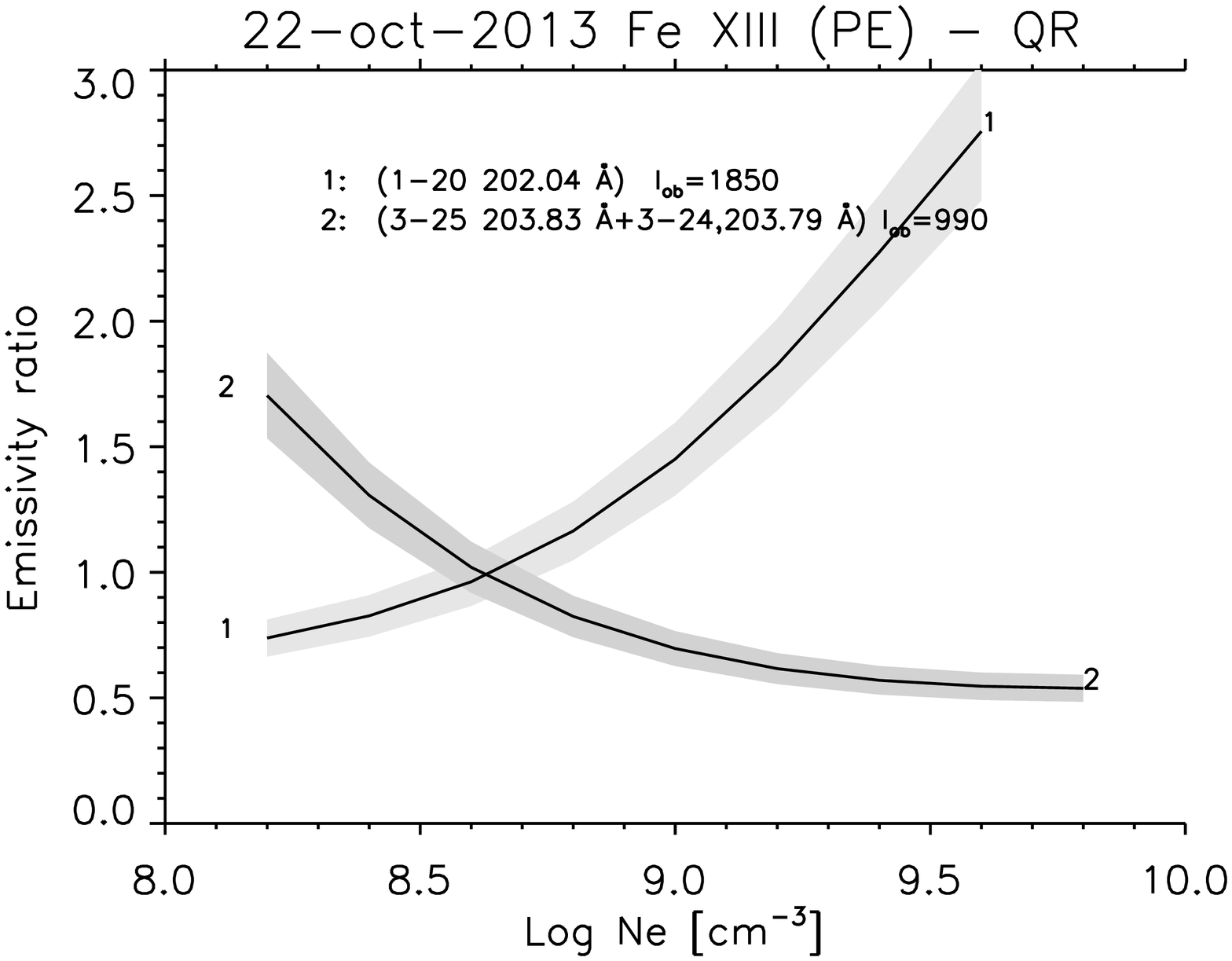}
\includegraphics[angle=-90,width=.9\hsize ]{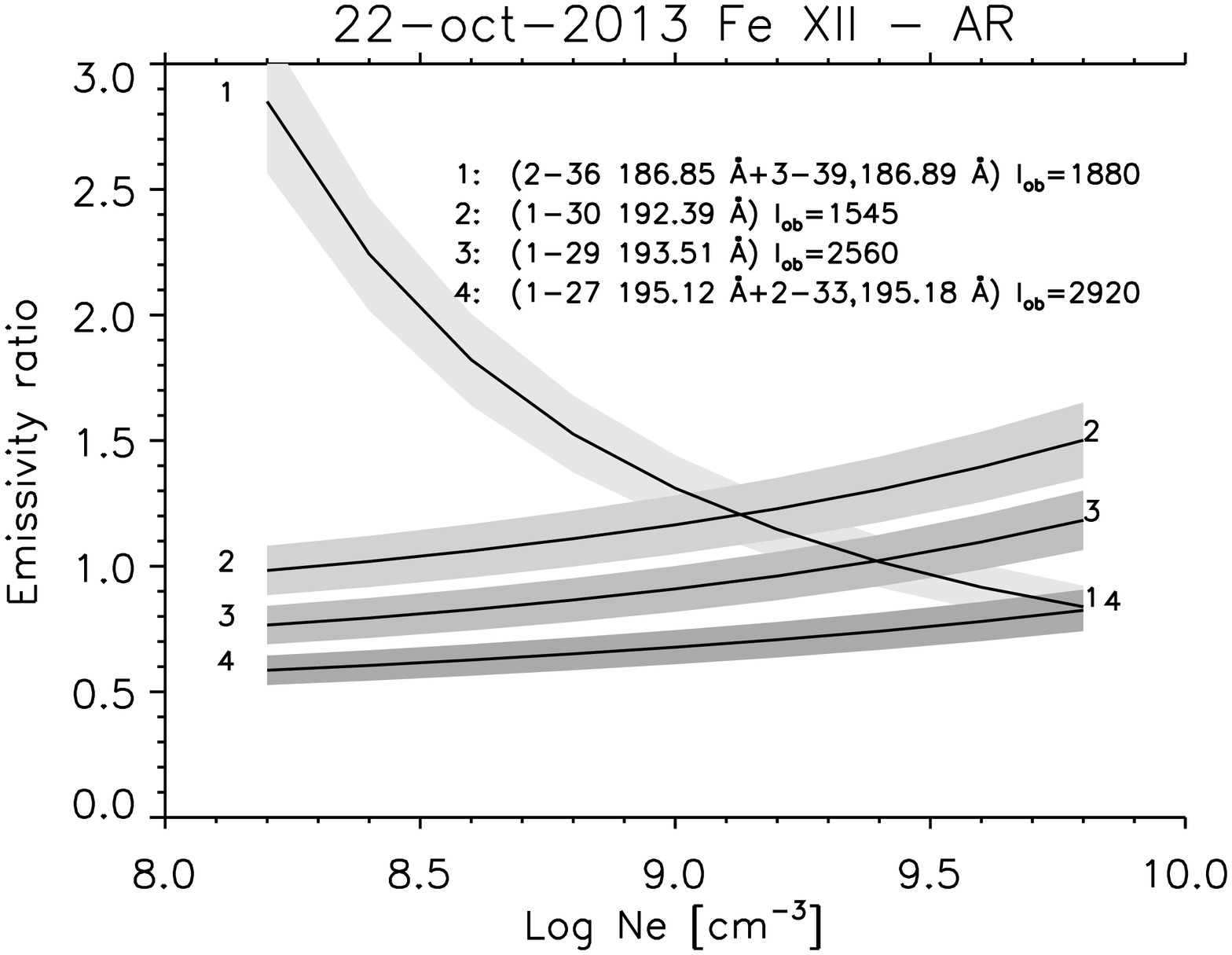}
\includegraphics[angle=-90,width=.9\hsize ]{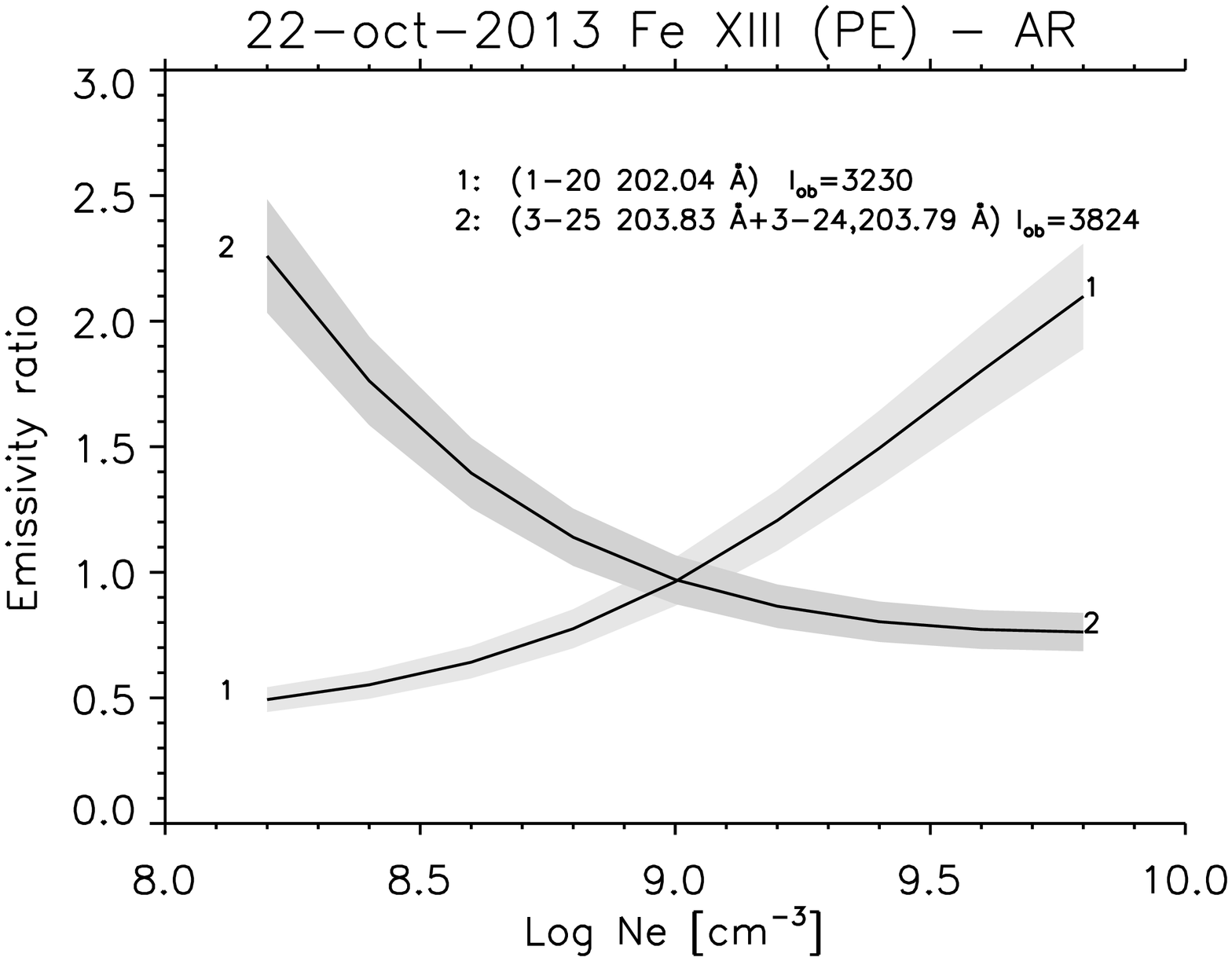}
\caption{Emissivity ratios of the EIS \ion{Fe}{12} and \ion{Fe}{13} lines, in the quiet off-limb region (above) and active region (below).}
\label{fig:fe_12_em}
\end{figure}

Figure~\ref{fig:fe_12_em} shows the emissivity ratios of the EIS \ion{Fe}{12} and \ion{Fe}{13} lines, in the quiet off-limb region (above) and active region (below).
It is clear that both regions are affected by opacity, which reduces the intensities of the \ion{Fe}{12} 193.5 and 195.1\,\AA\ lines, compared to the 192.4\,\AA\ one.
 The densities obtained from the \ion{Fe}{12} lines are close to those obtained from the \ion{Fe}{13} lines, considering the \ion{Fe}{12} 192.4\,\AA\ line, and the fact that this line is likely underestimated because of opacity effects (see discussion below).
We adopt the \ion{Fe}{13} densities as they are more reliable.
The QR and AR regions have densities around 4 and 10 $\times$ 
10$^8$ cm$^{-3}$. 

Note that the \ion{Fe}{13} lines include the photoexcitation effects, which affect the population of the ground state and the density diagnostics by up to 10\%, as discussed in \cite{dudik_etal:2021_eis_comp}.
They are caused by the large flux of photons emitted by the disk around 1\,$\mu$m, and resonantly absorbed by the two near-infrared \ion{Fe}{13} lines within the ground configuration. We have also explored the effects due to photoexcitation in the  \ion{Fe}{12} model ion, considering that several transitions within the ground configuration fall in the visible and far UV, but we did not find significant changes. We used observations of the solar irradiance in the far UV and visible.

\begin{figure}
\centering
\includegraphics[angle=-90,width=1.\hsize ]{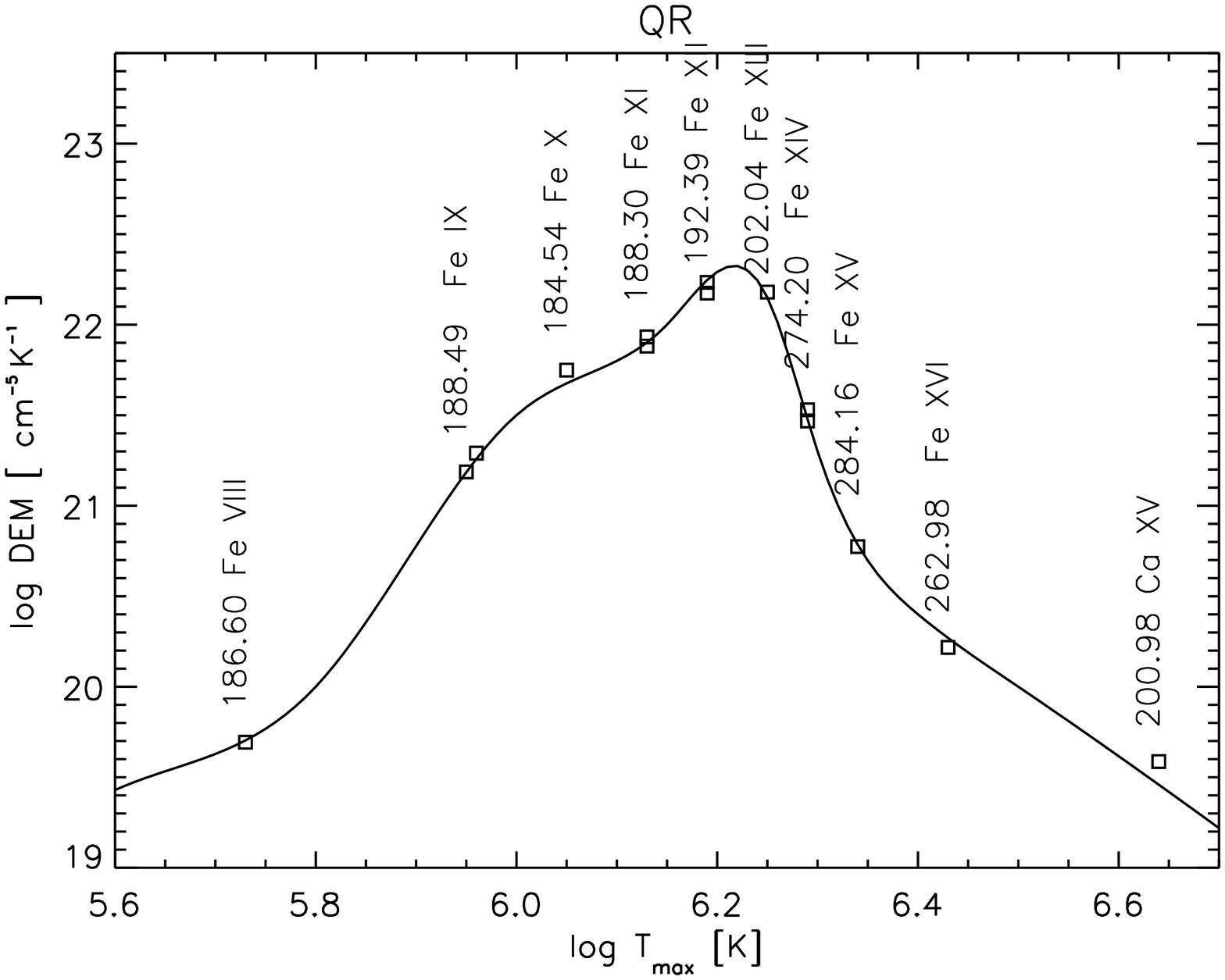}
\includegraphics[angle=-90,width=1.\hsize ]{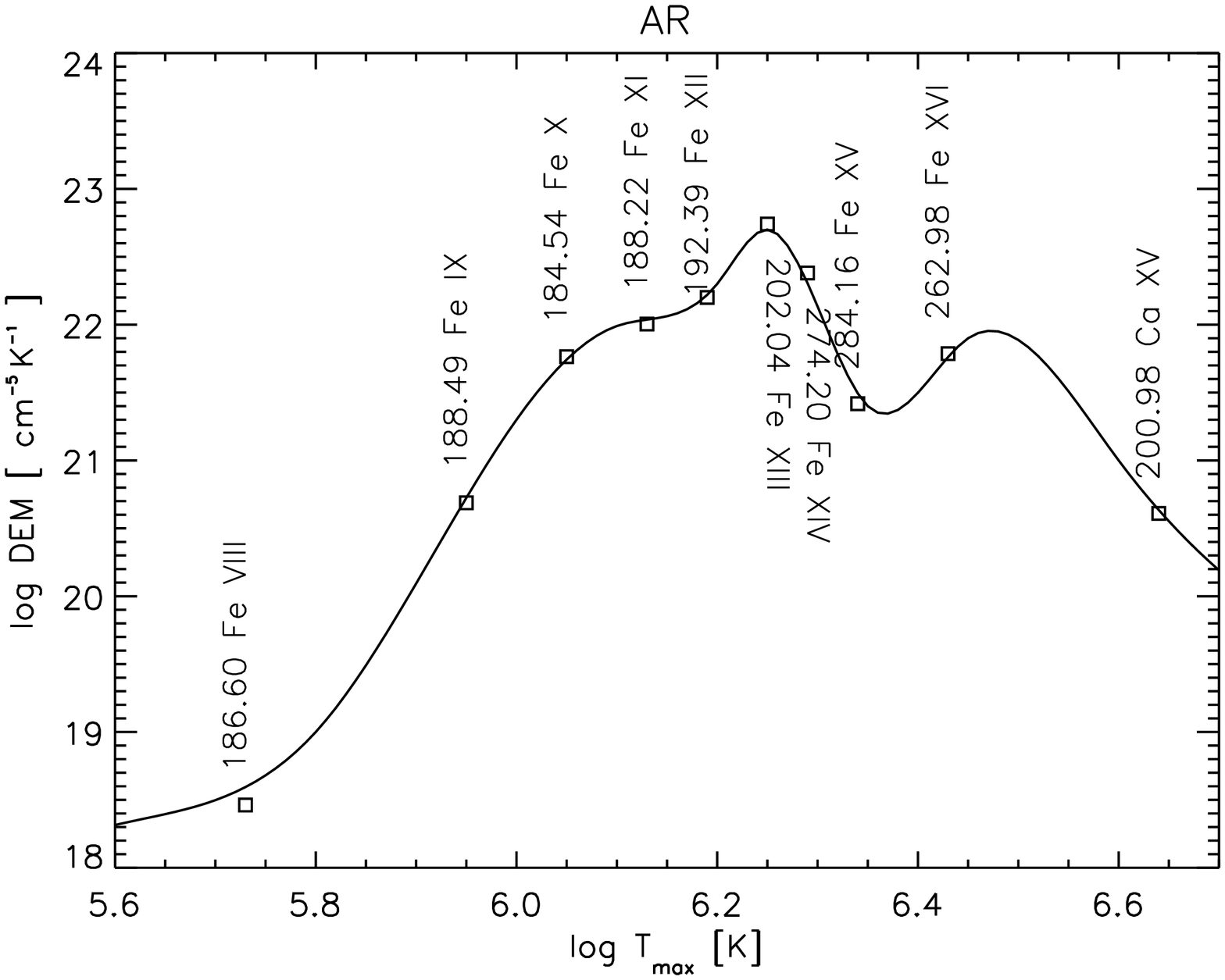}
\caption{DEMs for the quiet off-limb region
(above) and active region (below) for the 22-Oct-2013 observation. 
The points are plotted at the temperature $T_{\rm max}$ of the maximum in the emissivity, 
 and at the theoretical vs. the observed
 intensity ratio multiplied by the DEM value. The wavelengths (\AA) and main ion are indicated.}
\label{fig:dems}
\end{figure}

We have looked at the spatial distribution of various line ratios sensitive to temperature and found that the temperature so obtained is relatively constant in the off-limb regions. We 
produced EM loci plots for the quiet Sun and AR regions, finding that observations are consistent with an almost isothermal plasma around log $T$ [K]=6.2--6.25, which is the typical formation temperature of \ion{Fe}{12}. 
We have then performed a DEM analysis using a set of strong lines from Iron, not density-sensitive. The results are shown in Fig.~\ref{fig:dems} and confirm the near isothermality of the plasma emission, with a marked higher temperature component in the AR. The DEM analysis also indicated that the S/Fe relative abundance is close to photospheric around 1.2 MK (using a S X line). 

We regard the spatial variation in the 192.4\,\AA\,/\,1349.4\,\AA\ ratio as important, since this is independent of any calibration issues, and largely independent of the small variation in the density and temperature in the off-limb regions. The averaged ratio in the QR region (31) is close to the expected value, 34.1, obtained by folding the emissivities with the DEM distribution.
On the other hand, the AR value (21) is significantly lower than the expected value (30.9, with the DEM shown above). The lowest values near the limb (around 15) are even more difficult to explain. 

As there is no clear indication for absorption by filament material, 
and as opacity effects would decrease the 192.4\,\AA\ line by only a small amount (see Sect.~\ref{sec:tau}), we speculate that the main effect that could be responsible for changing the ratio is NMED. 
The fact that the ratio has  values close to the expected ones in the northern part of the off-limb region, suggests that  the EIS vs. IRIS  radiometric calibration is reasonably accurate.

\section{Possible effects on the \ion{Fe}{12} line ratio and the temperatures}
\label{sec:effects}

\subsection{Opacity effects}
\label{sec:tau}

Following \cite{delzanna_etal:2019_ntw}, the optical thickness at line centre can be written as
\begin{equation}
\tau_{0} = 8.3  \, 10^{-21} \, f_{lu}  \frac{\lambda^2}{\Delta \lambda_\mathrm{FWHM}} \; N_l \, \Delta S 
\end{equation}
where $f_{lu}$ is the absorption oscillator strength,
$N_l$ is the number density of the lower level, $\Delta S$ the path length,
$\Delta \lambda_\mathrm{FWHM}$ is the  FWHM of the line profile in \AA,
and $\lambda$ is the wavelength in \AA.
For the 195\,\AA\ line, $f_{lu}= 2.97/4$, neglecting the weaker line.

The  population of the lower level can be written as 
\begin{equation}
N_l = {N_l \over N({\rm Fe\, XII}) } \, {N({\rm Fe\, XII}) \over N({\rm Fe})} \,  Ab({\rm Fe})  \,
\frac{N_\mathrm{H}}{N_\mathrm{e}}  \, N_\mathrm{e} \;,
\end{equation}
where $N_l / N({\rm Fe\, XII})$  is the relative population of the ground state,
${N({\rm Fe\, XII}) / N({\rm Fe})}$  is the peak relative population of the ion,
$ Ab({\rm Fe})$ is the Fe  abundance, 
$N_\mathrm{H} /N_\mathrm{e} = 0.83$, and $N_\mathrm{e}$ is the averaged 
electron number density.

Considering the box above the active region, as we have assumed for photospheric abundances, we have $ Ab({\rm Fe}) = 3.16\, \times 10^{-5}$.
From the  EM loci / DEM analysis, we have $EM = 10^{28.3}$ [cm$^{-5}$] and log $T$[K]= 6.25,
approximately. With this temperature, ${N({\rm Fe\, XII}) / N({\rm Fe})}=0.21$
using the CHIANTI ionisation equilibrium.
Assuming the density from the \ion{Fe}{13} line ratio ($1 \times 10^{9}$ cm$^{-3}$,
we have $N_l / N({\rm Fe\, XII}) =0.75$ for these values of $T$, $N_\mathrm{e}$.
From the $EM$ and $N_\mathrm{e}$ values, assuming a filling factor of 1,
we obtain a path length of 2$\times$10$^{10}$ cm, from which
 we obtain $\tau_{0} = 0.96$ for the 195.1\,\AA\ line, and  
 $\tau_{0} = 0.32$ for the 192.4\,\AA\ line, as this transition
 has an oscillator strength a third of the 195.1\,\AA\ line.
 
Assuming that the source function $ S_\nu(\tau_\nu)$ does not vary along the line of sight,
 the peak intensity of each line is 
\beq
I_\nu =  S_\nu \, \left(1- e^{-\tau_0} \right) \;.
\eeq
Recalling that  the line source function $S_\nu$ is:
\beq
S_{\nu} = {2\,h\, \nu^3 \over c^2} \; \left( {g_u N_l \over g_l N_u} -1 \right)^{-1} \; ,
\eeq
with standard notation, we find that $S_{195} / S_{192} = 1.04$
using the statistical weights $g$ and the level populations calculated with the
model ion.

For $\tau_0 (195) =0.96$, the ratio of the intensities is then $I_{192} / I_{195}  = 0.43$, which is higher than the optically thin value of 0.31 and closer to the observed value of 0.53 for the region.

To estimate how much the weaker 192.4\,\AA\ line is suppressed for an optical depth of 0.32, as our simple assumption is equivalent to the  average escape factor formalism, we consider  the  homogeneous case  discussed by \cite{kastner_kastner:1990}, and obtain an escape factor of about 0.89, i.e., the 192.4\,\AA\ line is suppressed by about 10\%. Indeed if we increase the 192.4\,\AA\ line intensity by this amount, the emissivity ratio curves would result in a slightly lower  electron density, in better agreement with the values obtained from the  \ion{Fe}{13} ratio.

Finally, for the quiet off-limb `QR' region, if we repeat the above estimates,
 considering the lower $EM$ and lower density, we obtain
 $\tau_0 (192.4) = 0.33$, i.e., a similar optical depth, in agreement
 with the fact that the observed ratio is very similar.

%
\begin{figure}
\centering
\includegraphics[angle=0,width=0.8\hsize,viewport=0 10 495 380,clip]{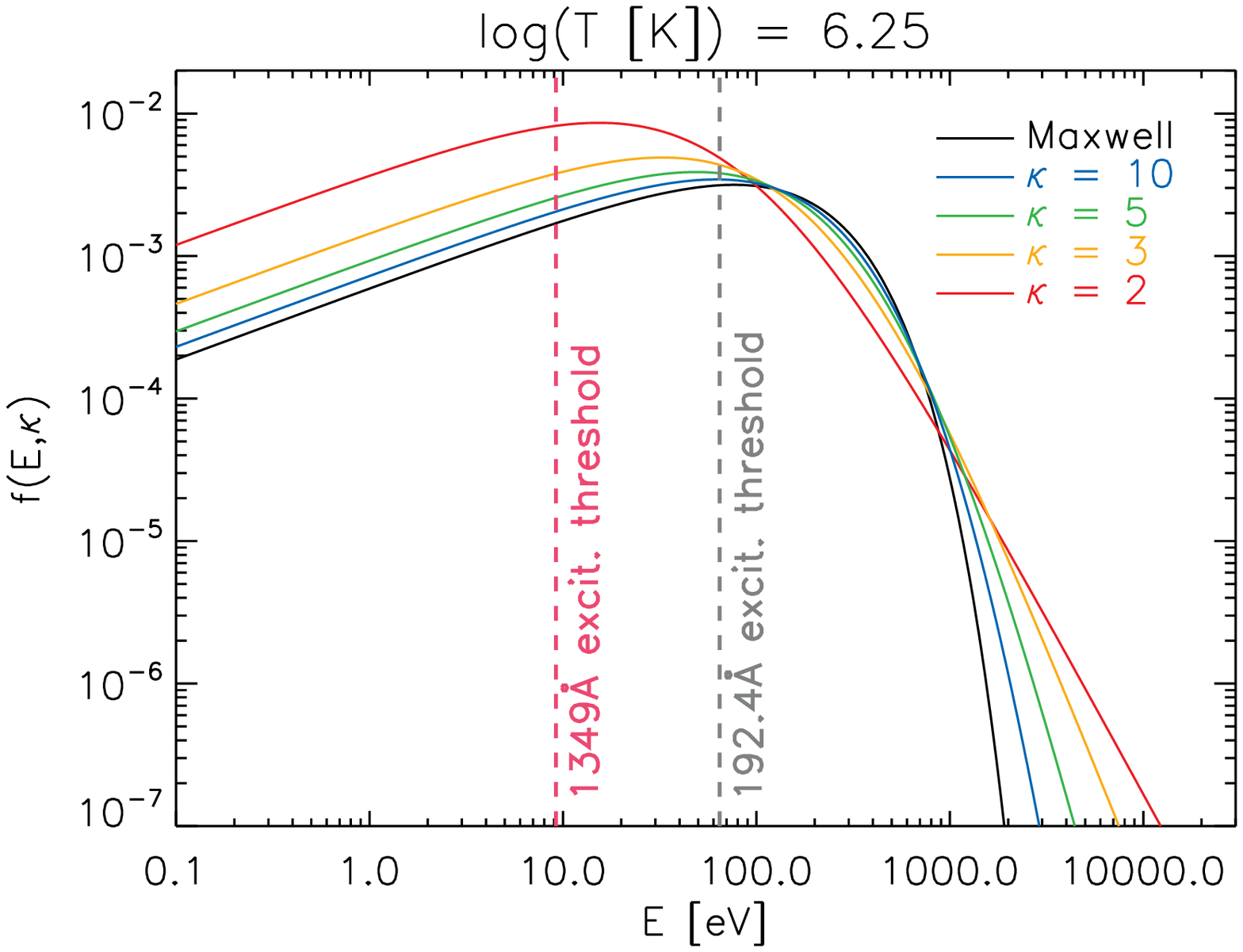}
\includegraphics[angle=0,width=0.8\hsize,viewport=0 50 495 380,clip]{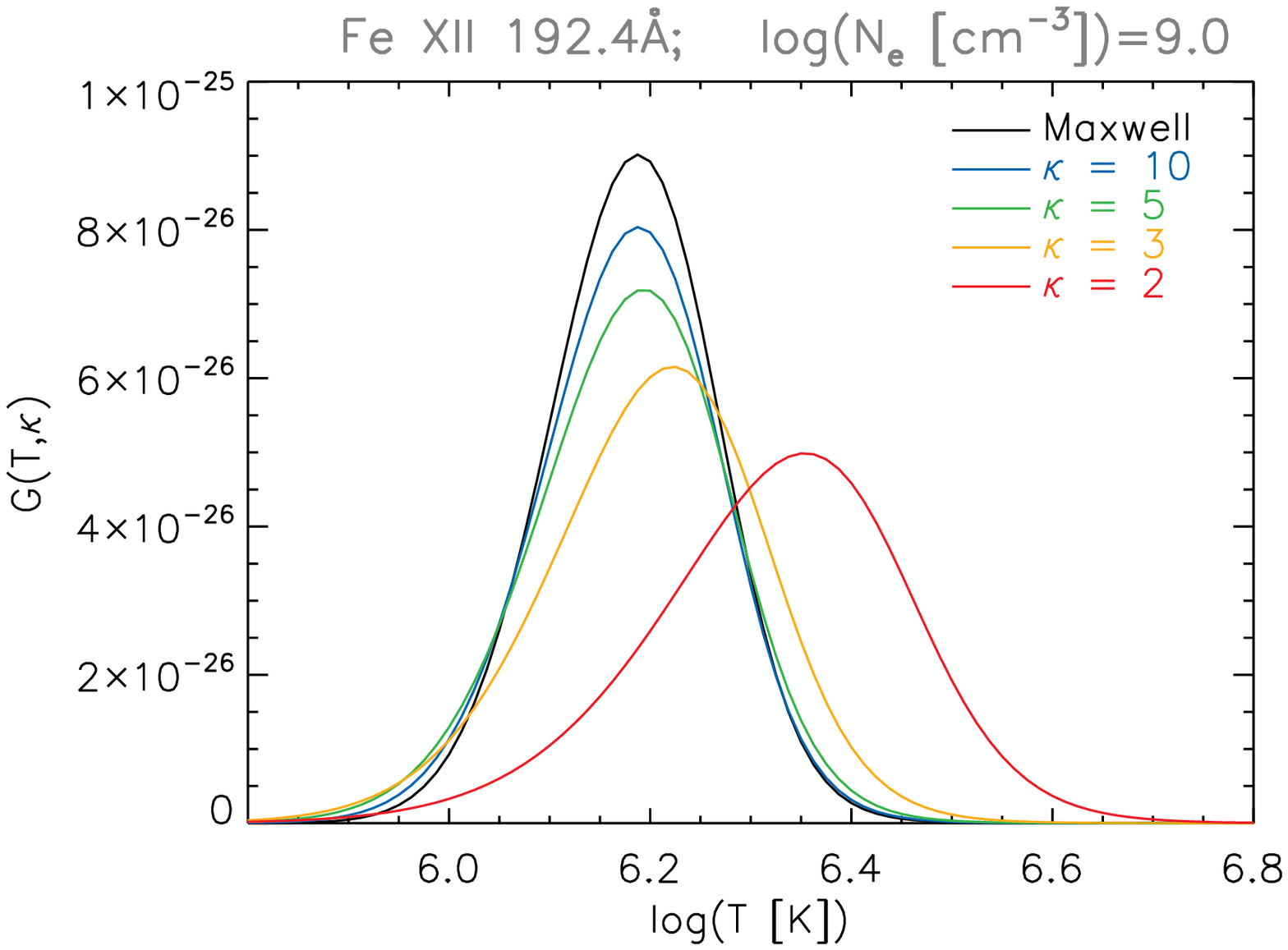}
\includegraphics[angle=0,width=0.8\hsize,viewport=0 0 495 380,clip]{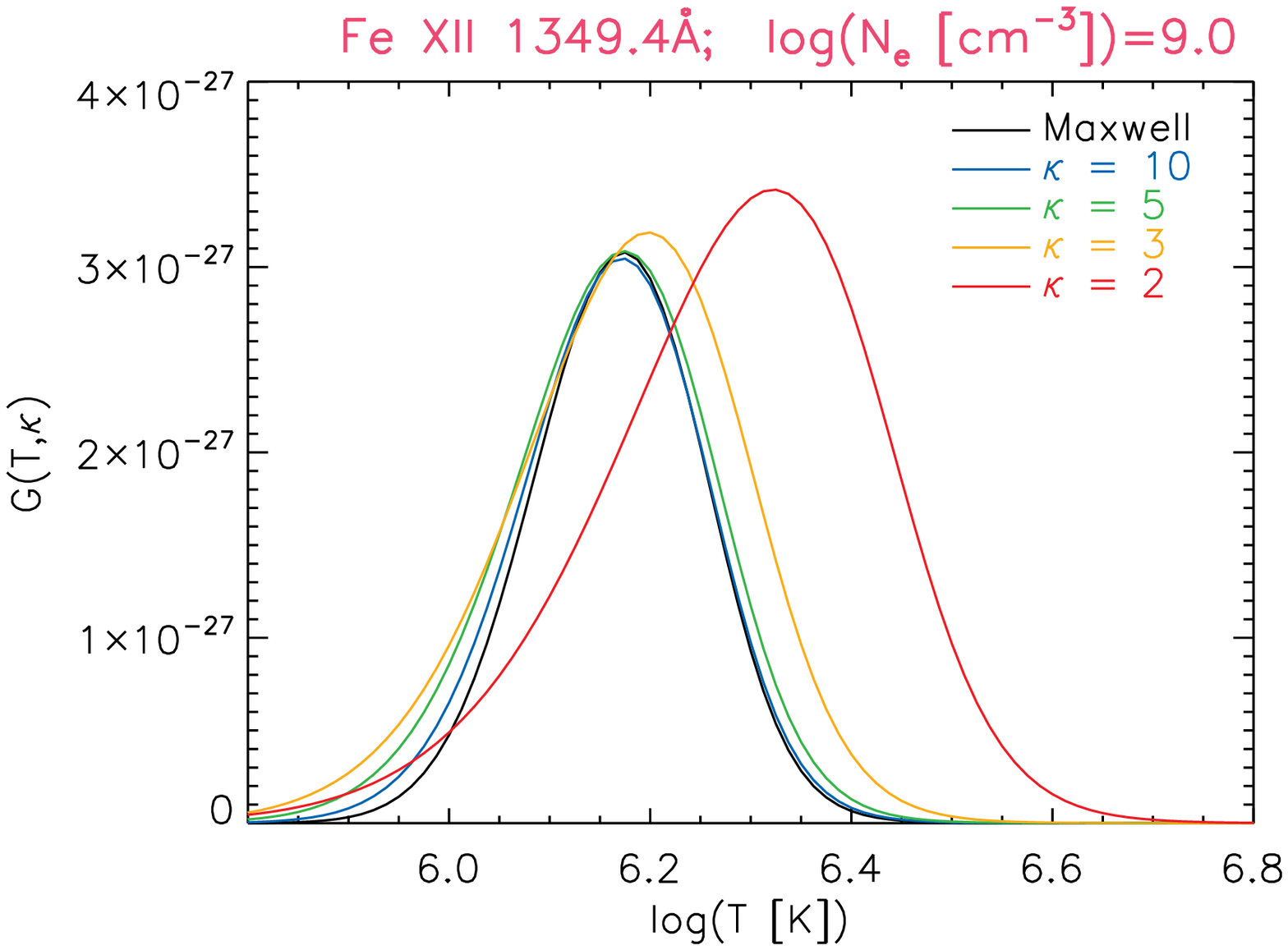}
\includegraphics[angle=0,width=0.8\hsize,viewport=0 10 495 380,clip]{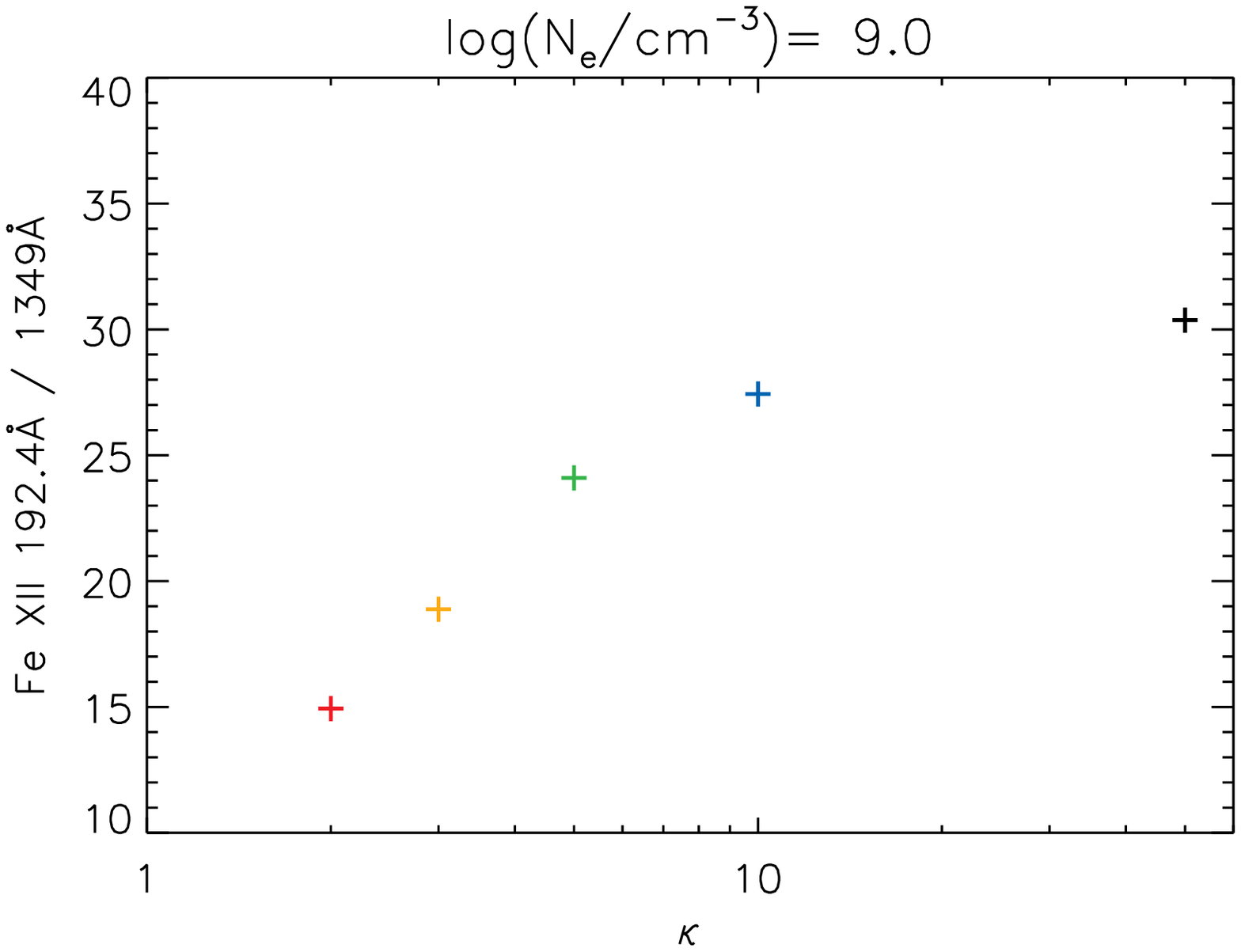}
\caption{Non-Maxwellian $\kappa$-distributions (top row) and their influence on the \ion{Fe}{12} 192.4\,\AA~and 1349.4\,\AA~lines, whose contribution functions are shown in the middle panels. The energy excitation thresholds for these two lines are denoted by dashed lines in the top panel. Bottom panel shows the behaviour of the 192.4\,\AA\,/\,1349\,\AA~ratio with $\kappa$, assuming peak formation temperatures.}
\label{fig:fe12_kappa}
\end{figure}
\begin{figure*}
\centering
\includegraphics[angle=0,width=0.49\hsize,viewport=20 0 498 498,clip]{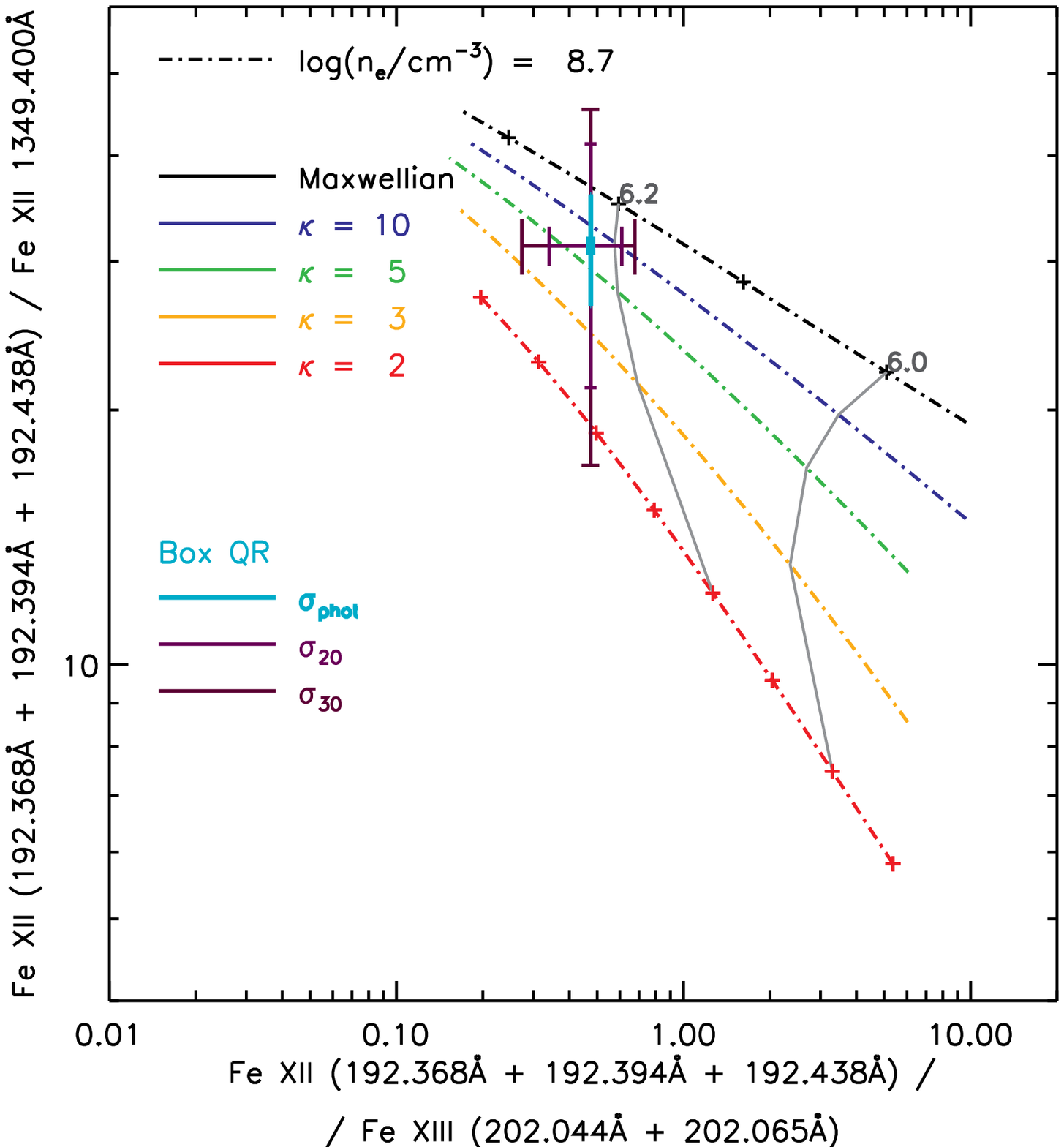}
\includegraphics[angle=0,width=0.49\hsize,viewport=20 0 498 498,clip]{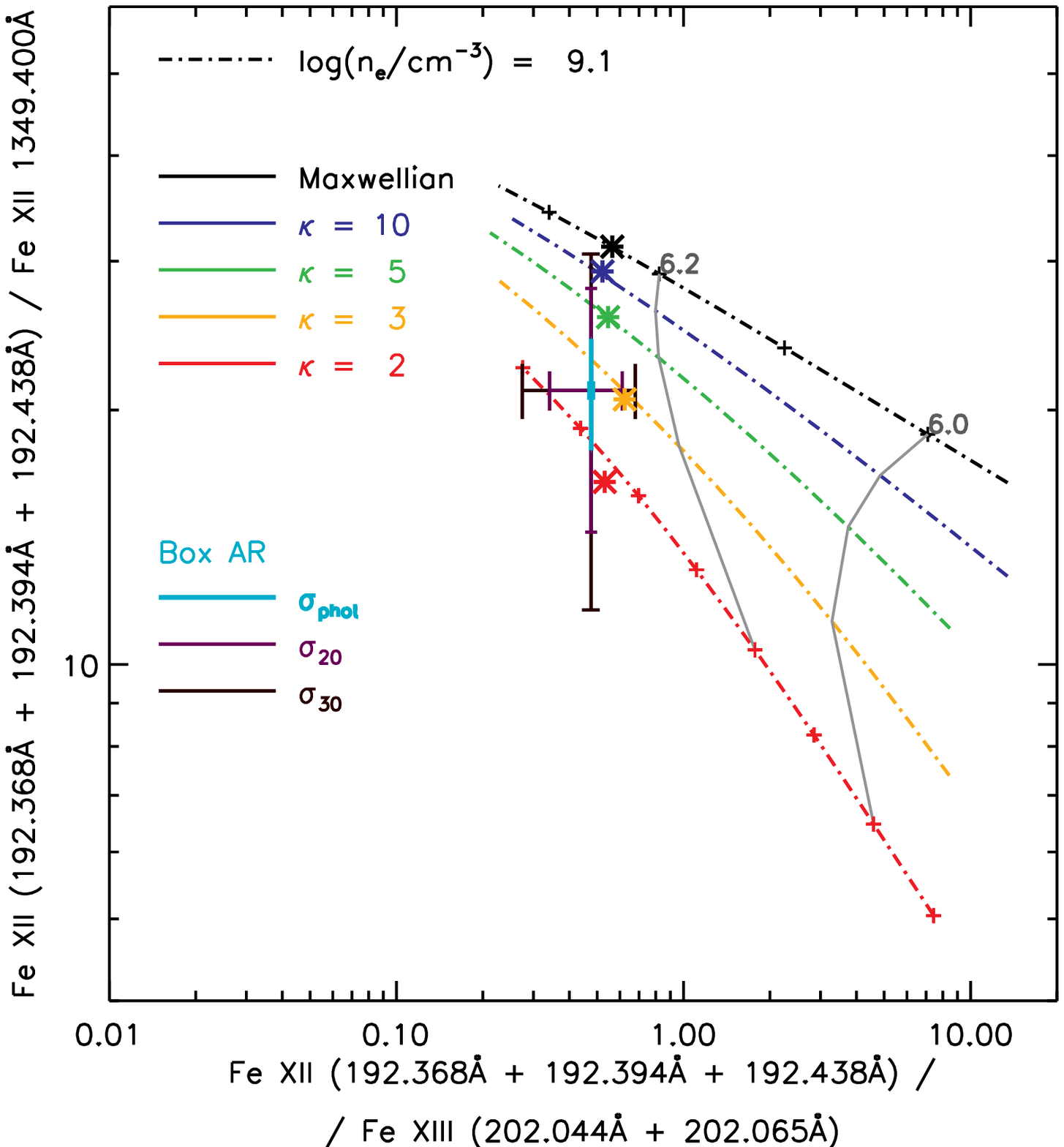}
\caption{Diagnostics of the NMED represented by $\kappa$-distributions using the ratio-ratio technique. Individual colors represent the value of $\kappa$, while the cross of different sizes represent the observed line ratios in the QR and AR boxes. The photon noise uncertainty $\sigma_\mathrm{phot}$ (light blue), as well as added 20\% to 30\% calibration uncertainties $\sigma_{20,30}$ (violet and black, respectively) are shown. Colored asterisks in the right panel denote the DEM$_\kappa$-predicted line intensity ratios (see Section \ref{sec:nmed_dems} for details). Note that both axes are scaled logarithmically.}
\label{fig:nmed}
\end{figure*}
\begin{figure*}
\centering
\includegraphics[angle=0,width=0.49\hsize,viewport=20 0 498 498,clip]{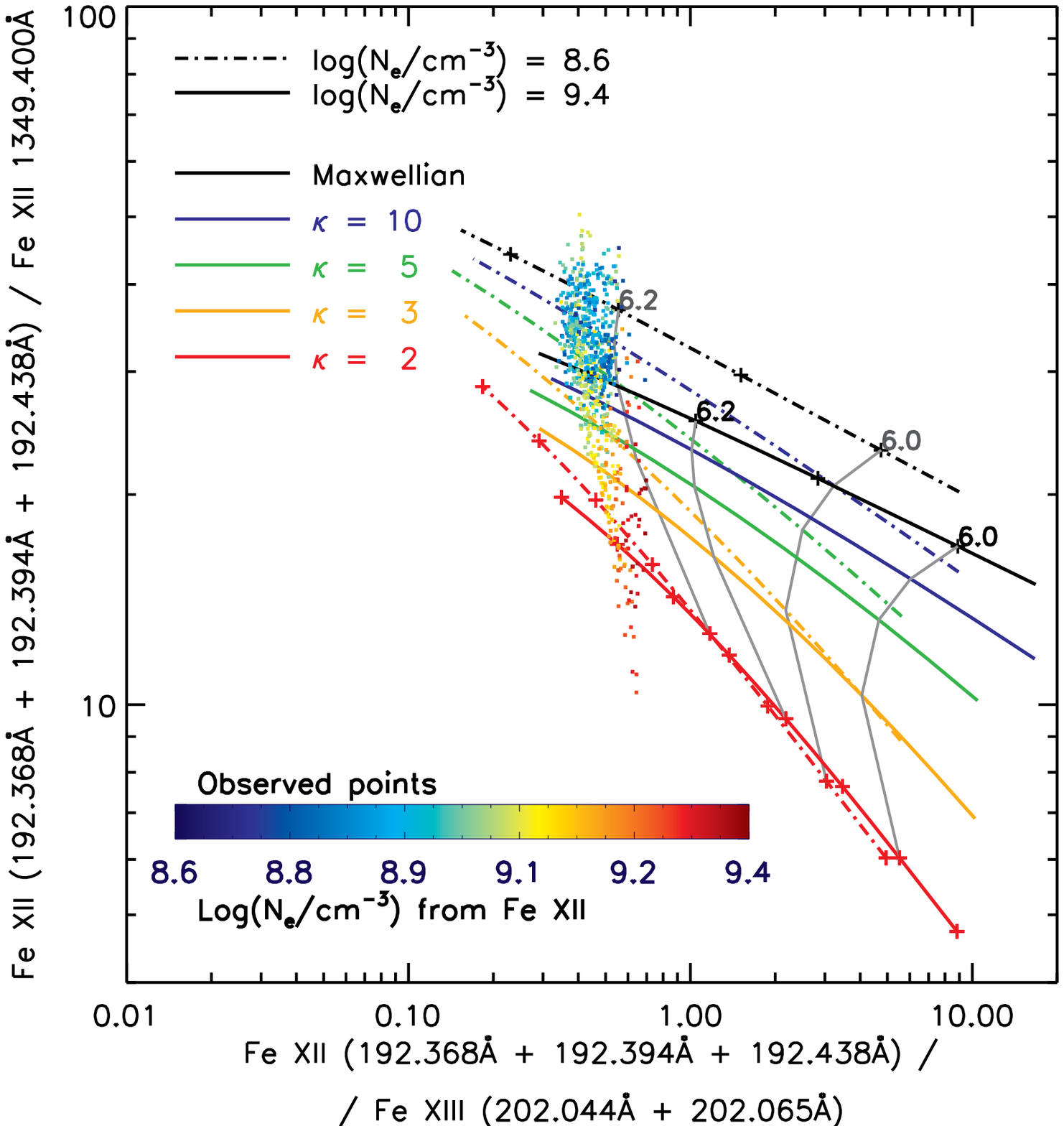}
\includegraphics[angle=0,width=0.49\hsize,viewport=20 0 498 498,clip]{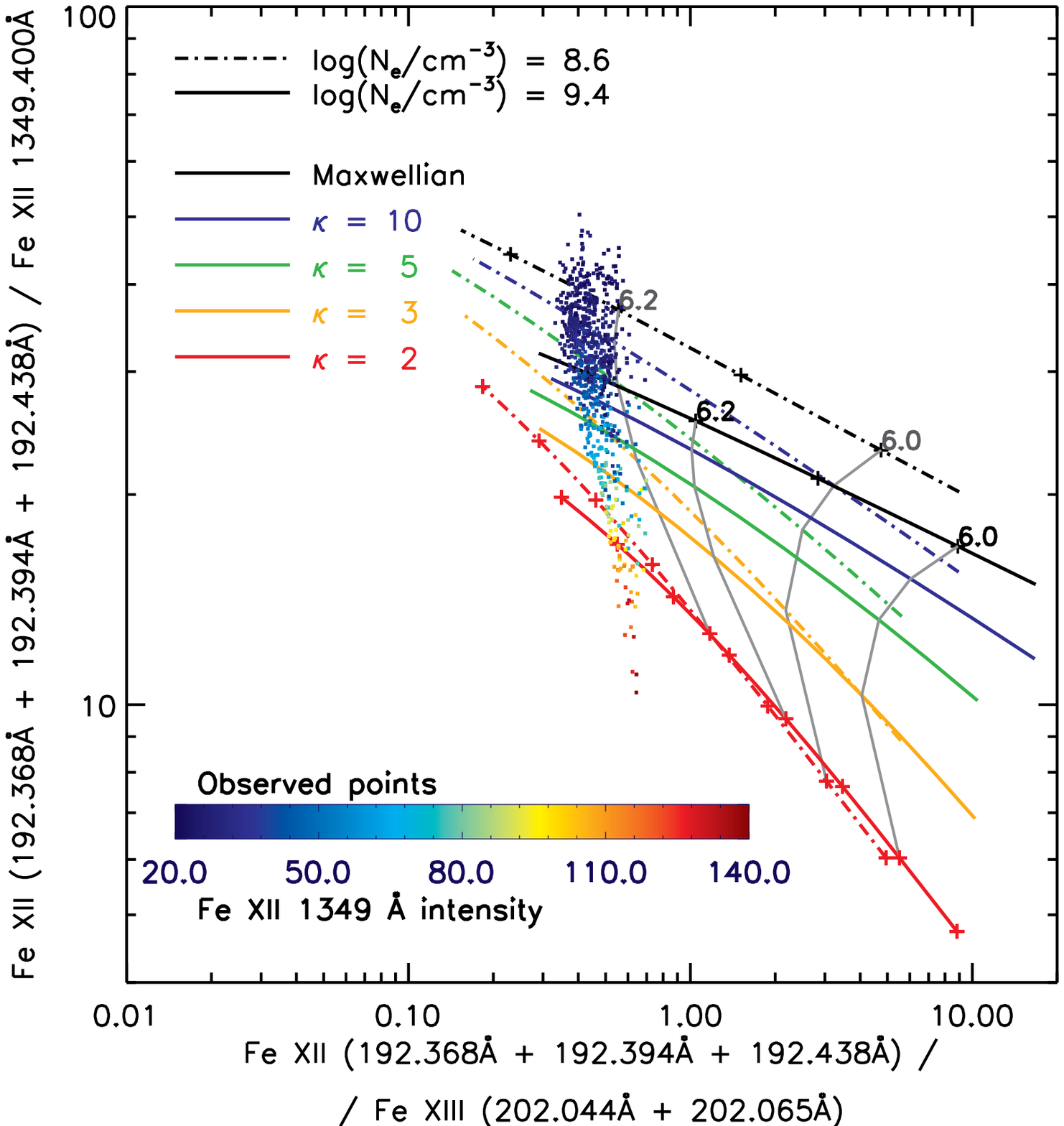}
\caption{Observed ratios in each individual pixel corresponding to Figure \ref{fig:eis_iris_ratio_sp2} are overplotted on the theoretical diagnostic curves. Two sets of curves are shown, for log($N_\mathrm{e}$\,[cm$^{-3}$])\,=\,8.6 and 9.4, representing the lowest and highest densities detected. The points are color-coded either according to the electron density (left panel) or according to the \ion{Fe}{12} 1349\,\AA~intensity (right panel). }
\label{fig:nmed_points}
\end{figure*}%
\subsection{Non-Maxwellian electron distributions (NMED)}
\label{sec:NMED}

\subsubsection{NMED effects on the \ion{Fe}{12} ratio}
\label{sec:NMED_effects}

To evaluate the effects of NMED, we considered the $\kappa$-distributions, a well known class of non-Maxwellian distributions characterized by a near-Maxwellian core and a power-law high-energy electron tail \citep[see, e.g.,][]{Livadiotis:2017,Lazar_Fichtner:2021}. We use the standard expression for $\kappa$-distributions of the second kind \citep[see the discussion in][]{dzifcakova:2021}, namely
\begin{equation}
	f_\kappa(E)dE= A_\kappa  \frac{2}{\sqrt{\pi}\left(k_\mathrm{B}T\right)^{3/2}} \frac{E^{1/2}dE}{\left(1+\frac{E}{(\kappa-3/2)k_\mathrm{B}T}\right)^{\kappa+1}}\,,
 \label{Eq:Kappa}
\end{equation}
where $E$ is the electron kinetic energy, $T$ is the temperature, $k_\mathrm{B}$ is the Boltzmann constant, and $A_\kappa$ is a constant for normalization to unity. From the expression above it follows that the slope of the high-energy power-law slope of the high-energy tail of a $\kappa$-distribution is $\kappa +1/2$.
The shape of the $\kappa$-distributions as a function of $E$ is depicted in the top row of Figure \ref{fig:fe12_kappa}.

The synthetic spectra for \ion{Fe}{12} and \ion{Fe}{13} were obtained using the KAPPA database \citep{dzifcakova:2015,dzifcakova:2021}, which allows for calculation of spectra for $\kappa$-distributions using the same atomic data as CHIANTI version 10 \citep{dere:1997,delzanna_etal:2021_chianti_v10}. We calculated the \ion{Fe}{12} and \ion{Fe}{13} line intensities for a range of temperatures $T$ and $\kappa$ values and found that the EIS/IRIS ratio of \ion{Fe}{12} 192.4\,\AA\,/\,1349\,\AA~line intensities offer unprecedented sensitivity to NMED, with the difference between Maxwellian and $\kappa$\,=\,2 being of about a factor of two, depending on temperature. This sensitivity to NMED comes from the widely different wavelengths, and thus excitation energy thresholds of the two lines - 192.4 and 1349\,\AA~\citep[cf.,][]{dudik_etal:2014_fe}. 

The line contribution functions $G(T,\kappa)$ of the two lines, equivalent to intensities normalized to unity emission measure, are shown in Figure \ref{fig:fe12_kappa}. For low $\kappa$\,=\,2, the peak formation of the \ion{Fe}{12} 192.4\,\AA~line occurs at higher $T$, and its intensity decreases. The shift in the temperature of the peak, as well as about half of the decrease of the peak, are due to the behaviour of the ionization equilibrium with $\kappa$ \citep{dzifcakova_dudik:2013,dzifcakova:2021}. The decrease in excitation due to relatively-lower amount of electrons in the $\kappa$\,=\,2 distribution at few hundred eV (top panel of Figure \ref{fig:fe12_kappa}) also contributes to the decrease of the peak of the \ion{Fe}{12} 192.4\,\AA~line. Compared to that, the forbidden 1349.4\,\AA~line intensity \textit{increases} for low $\kappa$ (bottom row of Figure \ref{fig:fe12_kappa}) despite the decrease of the relative ion abundance. The reason is chiefly that the forbidden line, whose excitation cross-section decreases with $E$, and which is excited by electrons at energies of $E$\,$\ge$\,9.2\,eV, experiences excess excitation by the relatively-higher peak of the $\kappa$\,=\,2 distribution (top row of Figure \ref{fig:fe12_kappa}).
The overall result is that for decreasing $\kappa$, the \ion{Fe}{12} 192.4\,\AA\,/\,1349\,\AA~line intensity ratio decreases  (bottom panel of Figure \ref{fig:fe12_kappa}).

However, one line ratio sensitive to $\kappa$ is not enough to determine the $\kappa$ from observations. This is because the distribution function has two independent parameters, namely $T$ and $\kappa$ (Equation \ref{Eq:Kappa}), which thus need to be determined simultaneously. \citep[e.g.,][]{dzifcakova_kulinova:2010,mackovjak_etal:2013,dudik_etal:2014_fe,dudik_etal:2015,lorincik_etal:2020,dzifcakova:2021}. Therefore, it is advantageous to combine this ratio with a primarily temperature-sensitive \ion{Fe}{12} / \ion{Fe}{13} ratio, which allows for de-coupling of the sensitivities to $\kappa$ and to $T$ (see Figure \ref{fig:nmed}) provided the plasma is in ionization equilibrium. For the latter ratio, we chose the \ion{Fe}{12} 192.4\,\AA~line together with the unblended and well-observed \ion{Fe}{13} 202.0\,\AA~line, thus minimizing the photon noise uncertainties.
The "ratio-ratio" diagnostic diagram for $T$ and $\kappa$ is then constructed by plotting the dependence on one line ratio upon the other one, see Figure \ref{fig:nmed}. There, the colored curves denote individual values of $\kappa$, with black being Maxwellian and red corresponding to $\kappa$\,=\,2. Individual values of log($T$\,[K]) are denoted by gray isotherms intersecting the curves for different $\kappa$. 

\subsubsection{NMED measurements}
\label{sec:nmed_measurements}

The line intensity ratios of \ion{Fe}{12} 192.4\,\AA~/ 1349\,\AA~together with the \ion{Fe}{12} 192.4\,\AA\,/\,\ion{Fe}{13} 202.0\,\AA~observed in the AR and QR boxes are shown in Figure \ref{fig:nmed} together with their uncertainties, consisting of photon noise uncertainty $\sigma_\mathrm{phot}$ (light blue) as well as the added 20--30\% calibration uncertainty, denoted as $\sigma_{20}$--$\sigma_{30}$ (violet and black crosses, respectively). This uncertainty is conservative, but is shown nevertheless because the instruments were not cross-calibrated independently. We note however that the differences in the observed \ion{Fe}{12} 192.4\,\AA~/ 1349\,\AA~ratio are systematic between the quiet Sun and active region (see Figure \ref{fig:eis_iris_ratio2}. That means the differences between AR and QR shown in the diagnostic diagram in Figure \ref{fig:nmed} are not a result of purely calibration uncertainty, since the calibration is the same for both the QR and AR. 
Note also that we have corrected the \ion{Fe}{12} 192.4\,\AA~line intensity for the optical depth effects, as discussed in Section \ref{sec:tau}.

In the QR box, where the observed ratio is higher and about 30, the plasma is consistent with the Maxwellian or weakly NME distribution within the uncertainties (left panel of Figure \ref{fig:nmed}). However, in the AR box, the observed ratio (of about 20) corresponds to NMED with the value of $\kappa$\,$<$\,5--10 even considering the calibration uncertainties. The value of $\kappa$ is possibly even lower, $\kappa$\,=\,2--3, as indicated by the photon noise uncertainty (Figure \ref{fig:nmed}).

We note that the theoretical diagnostic diagram consisting of the \ion{Fe}{12} 192.4\,\AA~/ 1349\,\AA~together with the \ion{Fe}{12} 192.4\,/ \ion{Fe}{13} 202.0\,\AA~line intensity ratios also show some dependence on electron density.
However, this dependence on density is much weaker than those of the  \ion{Fe}{11} line ratios previously employed for diagnostics of $\kappa$ by \citet{lorincik_etal:2020}. Given that electron density can be determined nearly independently of $\kappa$ \citep[see, e.g.,][and references therein]{dudik_etal:2014_fe}, we are confident that the current determination of NMED effects is not influenced by uncertainties in the determination of $N_\mathrm{e}$. 

The estimate of the uncertainty in electron density of $\approx$0.1 dex in log($N_\mathrm{e}$\,[cm$^{-3}$]) (see Figure \ref{fig:fe_12_em}) leads only to small changes in the theoretical diagnostic curves in Figure \ref{fig:nmed} (see Appendix \ref{Appendix:dens_tk}); meaning that the result of $\kappa$\,$\lesssim$\,5--10 in the AR box holds even when this uncertainty in the electron density is taken into account.

To illustrate the spatial variations in the NMED, we overplotted the ratios in all the pixels in the off-limb observation of 2013 October 22, corresponding to Figure \ref{fig:eis_iris_ratio_sp2}, on the NMED ratio-ratio diagrams (see Figure \ref{fig:nmed_points}). We color-coded the individual points either by the electron density $N_\mathrm{e}$ (left panel) or the observed \ion{Fe}{12} 1349\,\AA~intensity (right panel). The electron densities were measured using the \ion{Fe}{12} 186.9 / 192.4\,\AA~density-sensitive ratio, and were found to range between log($N_\mathrm{e}$\,[cm$^{-3}$])\,=\,8.6 to 9.4. We note that the highest values are found in the active region where the \ion{Fe}{12} 1349\,\AA~line is brightest. 

Figure \ref{fig:nmed_points} shows that the spread in the location of the observed \ion{Fe}{12} 192.4\,\AA\,/\,1349\,\AA~ratio is well matched by the theoretical curves. In agreement with Figure \ref{fig:eis_iris_ratio_sp2}, the larger \ion{Fe}{12} 1349\,\AA~intensities correspond to locations that are more non-Maxwellian. Finally, the \ion{Fe}{12} 192.4\,\AA\,/~\ion{Fe}{13} 202.0\,\AA~ratio, plotted on the horizontal axis, which is dominantly sensitive to $T$, indicates that the plasma is nearly isothermal, with all the points being clustered close to the log($T$\,[K])\,=\,6.25 isotherm. 

We therefore conclude that the NMED effects provide a possible explanation for the observed anomalously low  \ion{Fe}{12} 192.4\,\AA/1349\,\AA~line intensity ratios. 

\subsection{Plasma multithermality}
\label{sec:nmed_dems}

The diagnostics of $\kappa$ in the previous section assumed that the plasma can be described by two parameters, $\kappa$ and $T$. However, as we have seen earlier in Section \ref{sec:tau}, if interpreted as Maxwellian, the observations indicate presence of some degree of multi-thermality (see Figure 
\ref{fig:dems}). 

Generally, if the plasma is multi-thermal, the differential emission measure (DEMs) can be a function of $\kappa$ \citep{mackovjak_etal:2014,dudik_etal:2015,lorincik_etal:2020}. This has consequences for the diagnostics of $\kappa$, as these DEM$_
\kappa(T)$ could affect the predicted line intensities and their ratios that need to be compared with the observed ones. In fact, once the synthetic line intensities and their ratios are obtained for the respective DEM$_\kappa(T)$, each of the ratio-ratio diagnostic curves in Figure \ref{fig:nmed} collapses to a single point representing the two synthetic line intensity ratios predicted by the respective DEM$_\kappa$.

In order to take the possible plasma multithermality into account, we performed the DEM$_\kappa(T)$ inversions in the AR box for each $\kappa$ using the same method as in Section \ref{sec:tau}. In doing so, we used the respective line contribution functions $G(T,\kappa)$ as inputs. We note that this DEM analysis for variable $\kappa$ was done only for the AR box, as the quiet-Sun region (QR) intensities are already consistent with Maxwellian.

The DEM$_\kappa$-predicted points for each $\kappa$ are shown in the right panel of Figure \ref{fig:nmed} as series of colored asterisks, where the color represents the value of $\kappa$. It is seen that each point is close to the respective curve for the same $\kappa$, as expected. This analysis confirms that the \ion{Fe}{12} intensities in the active region can be explained by non-Maxwellian $\kappa$-distributions, as the points for $\kappa$\,=\,2--5 are a relatively close match to the observed intensities, while the Maxwellian point is still outside of the error-bars even if the calibration uncertainty is conservatively assumed to be 30\%.

\subsection{Time-dependent ionization (TDI)}
\label{sec:TDI}
In the presence of heating and cooling events occurring on short timescales, the possible effects of time-dependent ionization (TDI) on our diagnostics should also be considered. A full treatment of TDI requires detailed modelling of dynamic heating events in ARs, including its effect on both ion charge state distribution and the relative level population. As such, it is outside the scope of this work. Nevertheless we refer the reader to existing literature on this subject as well as theoretical arguments which indicate (to demonstrate) that TDI effects are likely not significant enough to explain the observed discrepancies in the intensity ratio of the two \ion{Fe}{12} lines. For instance, a relevant recent work is that of \citet{Olluri:2015}, who presented simulations of a quiet solar region from the three-dimensional magnetohydrodynamic code (MHD) code Bifrost \citep{Gudiksen2011} including non-equilibrium ionization, showing that the \ion{Fe}{12} ion was found to be close to its ionization equilibrium. Although a quiet Sun case might not be entirely applicable to our observations (the \fexii\ emission in a quiet region will be  primarily emitted in the corona whereas in a bright AR it will mostly be confined to the TR), we note that in the same simulation the TR ions were significantly out of equilibrium \citep[see Figure 15 in][]{Olluri:2015}. Another example comes from the simulations of nanoflare-heated coronal loops by \cite{bradshaw_klimchuk:2011}, where the ``warm'' emission, which includes \ion{Fe}{12} and \ion{Fe}{13}, was mostly close to equilibrium, even if the hotter emission was significantly out of equilibrium.

In the following paragraphs we also discuss possible effects of TDI on both on the (1) \ion{Fe}{12} relative emission as well as the (2) ion charge state distribution.

 (1)  TDI effects could lead to changes in the relative level population of \ion{Fe}{12}, and thus changes in the 192.4\,\AA\,/\,1349\,\AA~line intensity ratio.
 The EIS 192.4\,\AA\ line is an allowed transition with a very short
 decay time, of the order of picoseconds. 
 On the other hand, the IRIS  1349\,\AA\ forbidden line  is a decay from 
 the $^2$P$_{1/2}$ state, one of the 
 metastable levels in the ground configuration, which have typical decay times that are much longer. The lifetime of the $^2$P$_{1/2}$ level is only 4 milliseconds, so timescales this short would be needed to alter significantly the intensity of the IRIS line, compared to the equilibrium calculations. However, unlike the upper state of the EIS 192.4\,\AA\ line, which is solely populated from the ground state, the population of the $^2$P$_{1/2}$ is more complex. 
 To assess it, we have looked at the dominant processes, calculated in equilibrium at the temperatures and densities of the active regions we have observed. We find that about half of the  population of the $^2$P$_{1/2}$  is due to cascading from higher states, most of which are connected to the ground state, $^4$S$_{3/2}$. Nearly 30\% of its population comes from the  ground state, and  nearly 20\% from 
the $^2$D$_{5/2}$ state, which has a longer lifetime of 0.4 s. In turn, about 90\% of the $^2$D$_{5/2}$
population comes from cascading from high-lying states, which again are mostly connected to the ground state. 

Therefore, non-equilibrium effects with timescales shorter than 0.4 s would affect the population of the 
$^2$D$_{5/2}$ state but in turn change only by a small amount 
the intensity of the IRIS line. Overall, the ratio of the IRIS and EIS lines would be affected by at most 20\% if the timescales are shorter than 0.4 s.

(2) TDI effects could affect our observed ratios  through the ion charge distributions. The timescales for ion charge distributions to reach equilibrium are considerably longer in the solar corona. For example, at coronal densities, the \ion{Fe}{12} has an ionization equilibration timescale of the order of 10$^2$\,s \citep{smith_hughes:2010}, which is apt to be prolonged if there are flows in the plasma {that lead to mixing of plasma from regions of different temperatures.} Therefore, the TDI effects could affect the ionisation temperatures we have estimated. 
We recall that we estimated the temperature (via DEM analysis or line ratios) using lines from successive ionization stages of Iron. In particular, we used the  \ion{Fe}{12} / \ion{Fe}{13} line intensity ratio for simultaneous diagnostics of $T$ and $\kappa$ (see Sect. \ref{sec:NMED}). For the measured \ion{Fe}{12} 192.4\,\AA\,/\,1349\,\AA~ratio in the AR box to be consistent with Maxwellian, the complementary \ion{Fe}{12} 192.4\,\AA~/ \ion{Fe}{13} 202.0\,\AA~ratio would need to be different by about a factor of 10 (see the right panel of Figure \ref{fig:nmed}). 
This means that for the plasma to be Maxwellian, the \ion{Fe}{12} / \ion{Fe}{13} ratio should be at least 5 instead of the measured value of 0.5. Therefore, to explain the observations, the TDI effects would have to lead to departures from the \ion{Fe}{12} / \ion{Fe}{13} ratios by about at least an order of magnitude (cf. Figure \ref{fig:nmed}), which we deem unlikely, as the two ions are typically formed at similar temperatures and regions even in cases where the heating is transient and strong \citep[see, e.g., Figures 2--3 of][]{reale_orlando:2008}. 

Based on the considerations above, we suggest that TDI alone cannot easily explain the observed \ion{Fe}{12} ratios in our AR observations, although future numerical investigation will be necessary to rule it out completely.

\section{Discussion}
\label{sec:Discussion}

As described in Section \ref{sec:effects}, assuming that NMED are present  offers by itself a satisfactory explanation for the departures in the \ion{Fe}{12} 192.4\,\AA\, /\,1349\,\AA~line intensity ratio in the observed active regions. We now discuss the implications this finding entails, with emphasis on the timescales involved. These include:
\begin{itemize}
    \item timescale for equilibration of free electrons to a Maxwellian fluid,
    \item timescales for spontaneous emission,
    \item timescales for TDI effects,
    \item typical timescales for evolution of the AR emission,
    \item spectrometer exposure times,

    \item possible coronal heating frequency.
\end{itemize}
With the timescales for spontaneous emission and TDI effects were already discussed in Section \ref{sec:TDI}, we now examine the remaining ones, as well as their possible interplay.

\subsection{Timescales for maintaining NMED}
\label{sec:NMED_timescales}

Our analysis of the NMED effects was based on the $\kappa$-distributions (Section \ref{sec:NMED}), which have only one extra parameter, $\kappa$, and are assumed to be time-independent. However, once accelerated and non-Maxwellian, the bulk of the free electrons tends to thermalize due to collisions. Meanwhile, the same free electrons drive ionization, recombination, and excitation processes necessary for creation of the observed spectra. The timescale $\tau_\mathrm{e}$ for equilibration of the free electrons to a Maxwellian electron fluid {due to both electron--electron and electron--ion collisions} is given by Equation (3.50) of \citet{goedbloed_poedts:2004}, which in cgs units is:
\begin{equation}
   \tau_\mathrm{e} = \frac{1.09 \times10^{10}}{\mathrm{ln}\Lambda}\frac{\tilde T^{3/2}}{Z N_\mathrm{e}}\,,
    \label{Eq:tau_e}
\end{equation}

where ln$\Lambda$ is the Coulomb logarithm, $\tilde T$ is electron temperature in keV units, and $Z$ is the proton number. Taking $Z$\,=\,1 (considering that most of the ions in the solar corona are Hydrogen ions), ln$\Lambda$\,$\approx$\,10, and using the measured values of log($N_\mathrm{e}$\,[cm$^{-3}$])\,=\,9.1 and log($T$\,[K])\,=\,6.25 (corresponding to $\tilde T$ of about 0.22 keV), we obtain $\tau_\mathrm{e}$\,$\approx$\,0.1\,s.

We note that the above classical formula holds for the bulk of the electron distribution function, as the electrons in the high-energy tail are progressively less collisional, with the collision frequency decreasing with with kinetic energy $E$ as $E^{-3/2}$. In addition, the acceleration of progressively higher-$E$ electrons can also take longer \citep[see][]{bian_etal:2014}, although the details will depend on the acceleration mechanism itself; which, if indeed operating in the solar corona, is as of yet unknown. If the acceleration occurs due to turbulence, as derived by \citet{bian_etal:2014}, then the parameter $\kappa^*$\,=\,$\kappa+1$ describes the competing timescales of electron acceleration and collisional timescales, $\kappa^*$\,=\,$\tau_\mathrm{acc}/2\tau_\mathrm{coll}$ \citep[see Equation (14) of][]{bian_etal:2014}. It follows that if the measured $\kappa$ values as low as 2--3 in active regions are correct, the electrons must be continuously accelerated. Otherwise, we would not be able to see changes in the measured \ion{Fe}{12} 192.4\,\AA\,/\,1349\,\AA~ratio due to NMED effects, as the electrons would return to equilibrium Maxwellian distribution within a fraction of the exposure times required for our remote-sensing spectroscopic measurements.

In addition, it should be noted that the timescales for spontaneous emission in \ion{Fe}{12} (discussed in Section \ref{sec:TDI}) are much shorter, by orders of magnitude, than the electron equilibration timescale $\tau_\mathrm{e}$ derived above. Therefore, the level population of \ion{Fe}{12} reflects the changes in the electron distribution much faster, and is likely in equilibrium even in the case if the electron distribution undergoes evolution.

\subsection{Implications for coronal heating}
\label{Sec:Implicatons_heating}

{It is interesting to consider the implication of continuous re-acceleration of non-Maxwellian electrons (Section \ref{sec:NMED_timescales}) in terms of coronal heating.} We speculate that if continuous re-acceleration is connected to the frequency of the "nanoflare" heating of the solar corona, our observations may suggest  novel constraints on the nanoflare heating models.  We note that the current leading nanoflare or nanoflare train models \citep[see, for example,][and references therein]{Cargill:2014,Barnes:2016,Viall:2017,Reva:2018,Warren:2020} typically consider heating durations of the order of tens of seconds with separation between individual heating events as large as of the order of 10$^{2}$--10$^{3}$\,s. In addition, recent observations of moss variability in ARs with IRIS suggest that heating durations of the order of tens of seconds are common \citep{testa_etal:2013,testa_etal:2014,testa_etal:2020}.

Our implication that the re-acceleration occurs continously can be reconciled with these works if the heating occurs due to short individual bursts (so that electrons are re-accelerated), while the duration of the envelope of the heating can be as long as 10$^{1}$--10$^{2}$\,s. One mechanism that behaves this way is slipping reconnection, which is the general mode of reconnection in three dimensions \citep[see, e.g.][]{janvier_etal:2013,dudik_etal:2014_slipping}. During slipping reconnection, individual field lines reconnect many times, indeed sequentially, with different field lines, while their footpoints slip across the solar surface. The slipping reconnection in many small-scale quasi-separatrix layers has been shown to be a viable coronal heating mechanism \citep[][]{yang_etal:2018nat} and is indeed sometimes observed to occur in moss regions \citep[][]{testa_etal:2013}. However, other mechanisms can also lead to many individual heating events occurring due to a longer-duration conditions of energy release in a coronal loop. One can imagine that, for example, wave-particle resonance interactions would behave much the same way as long as the larger-scale wave lasts. Such speculations are however out of the scope of the present work, and we do not engage in them further. Nevertheless, we do note that if the  scenario of frequent re-acceleration events occurring within a longer-duration heating envelope is correct, the behavior of emission within individual emitting strands (as well as their collective emission) should be modeled in detail, as there are many timescales involved, as mentioned at the beginning of this section, including the timescale for equilibration of the relative level population, TDI effects, and the NMED effects.

\section{Summary}

We have investigated coordinated Hinode/EIS and IRIS observations of \ion{Fe}{12} lines. While the EIS observes the allowed lines in the EUV part of the spectrum, the IRIS observes the forbidden line at 1349\,\AA. We find that the ratio of these two lines decreases strongly with the increase in intensity of the forbidden 1349\,\AA~line in active regions. 
In the quiet Sun, the \ion{Fe}{12} 192.4\,\AA\,/\,1349\,\AA~ratio is about 30--40, while in active regions, the ratio decreases down to values of below 20, even reaching values as low as 10 in some cases. These measurements were accompanied by determination of the temperature and emission measure using lines of \ion{Fe}{9}--\ion{Fe}{16}, as well as electron densities using density-sensitive \ion{Fe}{12} and \ion{Fe}{13} lines from EIS.

Using synthetic spectra obtained from CHIANTI version 10, we investigated whether the behaviour of the \ion{Fe}{12} 192.4\,\AA\,/\,1349\,\AA~ratio could be due to its dependence on electron temperature and density. Especially in active regions, we found significant and systematic discrepancies in the observed 192.4\,/\,1349\,\AA~ratio with respect to the predictions based on the synthetic spectra obtained by CHIANTI. In the AR box that we selected for detailed analysis, we measured values of log($T$\,[K])\,=\,6.25 and log($N_\mathrm{e}$\,[cm$^{-3}$])\,=\,9.1, resulting in a predicted \ion{Fe}{12} 192.4\,/\,1349\,\AA~ratio of about 30, while the observed value is about 20. 

We reviewed the potential causes of this discrepancy, including:
\begin{enumerate}
    \item Opacity effects on the \ion{Fe}{12} EUV lines,
    \item presence of cool plasma along the line of sight,
    \item plasma multithermality,
    \item dependence of the observed ratio on non-Maxwellian electron distributions (NMED).
    \item effects due to time-dependent ionization (TDI)
\end{enumerate}

Opacity in the \ion{Fe}{12} lines was detected as an increase in width of the EUV lines, especially the 193.5 and 195.1\,\AA~lines \citep[see][]{delzanna_etal:2019_ntw}. Being the weakest of the three transition, the 192.4\,\AA~line is least affected. Based upon the measured temperatures and emission measures, we estimated that the optical depth in the 192.4\,\AA~line is about 0.32 (and 0.96 for the 195.1\,\AA~line), leading to suppression of the \ion{Fe}{12} 192.4\,\AA~line by about 10\%. This effect was therefore deemed insufficient to explain the discrepancies in the 192.4\,\AA\,/\,1349\,\AA~ratio. We subsequently corrected the observed 192.4\,\AA~line intensity accordingly to account for self-absorption.

The relative absence of cool material along the line of sight was checked based on the AIA 193\,\AA~and H$\alpha$ observations by the Kanzelh\"ohe Solar Observatory. We note that the two wavelengths have similar optical thickness \citep{anzer_heinzel:2005} and that the absorption near 195\,\AA~occurs due to the \ion{H}{1}, \ion{He}{1}, and \ion{He}{2} continua. Our selected QR and AR for the quiet and active region were also chosen to be above the H$\alpha$ spicules, and in regions devoid of prominence material, so that the absorption by the H and He continua was deemed negligible. 

We used the the $\kappa$-distributions to study the influence of the NMED on the line ratio. Using the updated KAPPA database \citep{dzifcakova:2021} corresponding to CHIANTI version 10, it was found that the \ion{Fe}{12} 192.4\,\AA\,/\,1349\,\AA~ratio decreases with increasing number of high-energy electrons (i.e., lower $\kappa$). The observed \ion{Fe}{12} ratio of about 20 in the AR can be explained by NMED with $\kappa$ as low as 2--3, although calibration uncertainties are significant. In addition, the spatial distribution of the ratio matches well the theoretical diagnostic curves for NMED, where the lowest observed ratios correspond to strongly NMED plasmas. These theoretical curves for NMED are only weakly dependent on electron density and show strong sensitivity to $\kappa$, making the \ion{Fe}{12} ratio one of the best diagnostic options for the NMED. In addition, the plasma multithermality was ruled out as the cause of the departure of the \ion{Fe}{12} ratio in active regions, since any DEM effects would only exacerbate the the discrepancy. 

Finally, based on theoretical arguments as well as existing literature, we concluded that TDI effects alone are likely insufficient to explain the observed discrepancies in the \ion{Fe}{12} ratio, although they cannot be ruled out.

Our measurements employed a new EIS calibration, which will be described in detail in a separate publication. The uncertainty inherent in the calibration limits the determination of $\kappa$ from our measurements. Nevertheless, the off-limb quiet Sun and active region are observed simultaneously, and the new calibration shows that the ratio in the quiet Sun is consistent with Maxwellian electrons, in accordance with independent previous measurements from EIS \citep{lorincik_etal:2020}, but also X-ray instruments \citep[][]{kuhar_etal:2018}, which do not show presence of accelerated particles in quiet Sun regions. This indicates that the relative EIS/IRIS calibration is likely correct. 

For the reasons listed above, we are left with NMED as the most likely, simplest cause of the anomalously low \ion{Fe}{12} 192.4\,\AA\,/\,1349.4\,\AA\ ratio in the observed active regions.

{Using Equation (3.50) of \citet{goedbloed_poedts:2004} we calculated that the timescale $\tau_\mathrm{e}$ for equilibration of the free electrons to a Maxwellian electron fluid is given by $\tau_\mathrm{e}$\,$\approx$\,0.1\,s, for the core of the distribution, using the values of temperature and density measured.  Given that the \ion{Fe}{12} lines were observed with exposure times of tens of seconds, this suggest that the electrons must be continuously accelerated or re-accelerated over these timescales, otherwise they would return to equilibrium Maxwellian distribution within a fraction of second. Our observations could thus provide interesting new constraints on the nanoflare-based coronal heating models.}

Observations with well-calibrated instruments in the future could use these or similar allowed-to-forbidden coronal line ratios to diagnose the presence of NMED. One attractive option is EUVST, as it will observe the same lines as the EIS SW channel, and UV  lines with a high sensitivity, hopefully measuring the diagnostic ratios with a cadence of a fraction of a second.

\acknowledgments
 GDZ and HEM acknowledge support from STFC (UK) via the consolidated grants to the atomic astrophysics group (AAG) at DAMTP, University of Cambridge (ST/P000665/1. and ST/T000481/1). VP was supported by NASA under contract NNG09FA40C ({\it IRIS}).
PT was supported by contract 8100002705 from Lockheed-Martin to SAO, NASA contract NNM07AB07C to the Smithsonian Astrophysical Observatory, and NASA grant 80NSSC20K1272.
J.D. and E.Dz. acknowledge support from Grants No.\,20-07908S and 22-07155S of the Grant Agency of the Czech Republic, as well as institutional support RVO:67985815 from the Czech Academy of Sciences. GDZ, JD, and HEM also acknowledge support from the Royal Society via the Newton International Alumni programme.
We thank the anonymous referee for careful reading and useful comments.
IRIS is a NASA small explorer mission developed and operated by LMSAL with mission operations executed at NASA Ames Research center and major contributions to downlink communications funded by ESA and
the Norwegian Space Centre.
Hinode is a Japanese mission developed and launched by ISAS/JAXA, with NAOJ as a domestic partner and NASA and STFC (UK) as international partners. It is operated by these agencies in cooperation with the ESA and NSC (Norway).
SDO data were obtained courtesy of NASA/SDO and the AIA and HMI science teams. {H$\alpha$ data were provided by the Kanzelh\"{o}he Observatory, University of Graz, Austria.}

\bibliography{fe_12}
\bibliographystyle{aasjournal}

 \clearpage

 \appendix

\section{ Blending of the IRIS \ion{Fe}{12} line}

As the IRIS spectral range is rich in unidentified narrow lines
due to photospheric/chromospheric lines, as well as molecular
lines, we have analysed one full spectral IRIS atlas 
during a flare, and measured the intensities of all the
known and unknown cool lines in the spectra.
The observation was obtained on 2013-10-11 with a long
dense IRIS raster that started at 23:55 UT.

\begin{figure}
\centering
\includegraphics[angle=90,width=0.5\hsize ]{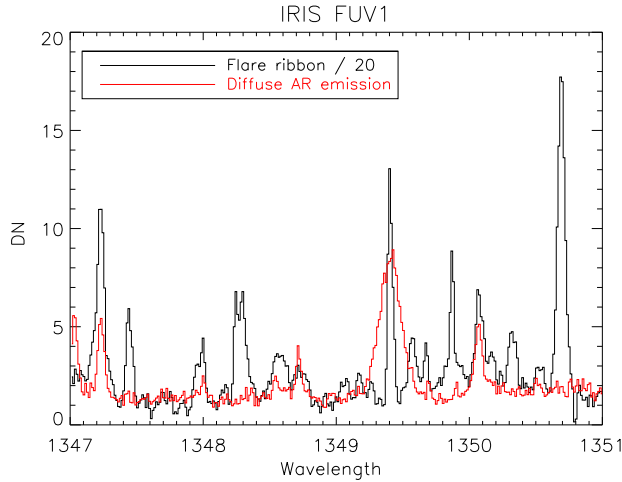}
\caption{IRIS FUV1 spectra around the \ion{Fe}{12} 1349.4\,\AA\ line. }
\label{fig:flare_ribbon}
\end{figure}

Fig.~\ref{fig:flare_ribbon} shows superimposed two spectra, one obtained
on a moss region (pixel values 240:260 in solar X and 130:160 along the slit),
and the other one at pixel coordinates 73 (solar X) and 192 (solar Y) on
a flare ribbon, reduced by a factor of 20. It is clear that in the moss
region the 1349.4\,\AA\ line is due to \ion{Fe}{12}, as it has the expected width.
The spectrum in the ribbon is instead solely due to an unidentified
narrow cool line, with a wavelength coincident with that of the
\ion{Fe}{12} line.
The spatial distribution of this unidentified line is quite different than
that of most known lines such as the  \ion{Cl}{1} 1351.66\,\AA.
It is similar to that of the  strong \ion{C}{1} lines at
1357.13, 1354.28\,\AA, but is actually  closest in morphology to
another unidentified line at 1350.69\,\AA.
The ratios of the known \ion{C}{1} lines are relatively constant,
so a possible way to estimate the contribution of the unidentified line
at 1349.4\,\AA\ is to consider the observed ratios  in the ribbons
with the \ion{C}{1} lines. For example the ratio (in data number)
with the \ion{C}{1}  1354.28\,\AA\  ranges between 0.02 and 0.07.

\section{EIS radiometric calibration}

Briefly, a DEM analysis was applied to off-limb quiet Sun observations close in time to the observations discussed here, to obtain  the relative EIS calibration using the strongest coronal lines. The advantage is that the plasma is nearly isothermal with an isodensity in these cases and possible issues related to the presence of NMED are avoided. Also, this removes blending with cool lines from a few coronal lines. This is an extension of the method used by \cite{warren_etal:2014}, where strict isothermality was assumed. 
 The established relative calibration for the short-wavelength (SW) channel was then used to calibrate the EIS spectra, for a direct cross-calibration with simultaneous
SDO AIA 193\,\AA\ data, taking into account the different spatio-temporal resolutions, basically following the methods described in \cite{delzanna_etal:2011_aia,delzanna:2013_multithermal}.
Good agreement (to within a few percent) between the AIA DN/s predicted from EIS, and those observed by AIA (re-scaled to the lower spatio-temporal
resolution of EIS) is imposed, noting that a  typical spatial scatter around 10\% is normally found. We used the modelled AIA degradation as available in SolarSoft, with the option of the normalisation with SDO EVE. 
We also  checked this  AIA calibration against simultaneous SDO EVE observations, using the latest EVE calibration, which in itself relies on a comparison with a few sounding rocket flights and adjustments using line ratios, following the methods adopted for EIS \citep{delzanna:13_eis_calib}. In turn, the prototype EVE flown on the sounding rocket flights is regularly calibrated on the ground. The absolute calibration of the EVE prototype is deemed accurate to within 20\%, although detailed comparisons carried out on the first flight showed
larger discrepancies (40\%) for some of the strongest lines
\citep{delzanna_andretta:2015,delzanna:2019_eve}. The  overall accuracy of the EIS absolute calibration adopted here, considering all the comparisons, could be estimated to be in the range 20--30\%. Such a reliable calibration in the EUV (a notoriously difficult problem) could only be established for EIS data in 2013 and 2014, as in 2014 the failure of EVE MEGS-A meant that no direct AIA/EVE cross-calibrations could be carried out. After 2014, the only useful cross-calibration EVE sounding rocket
was flown in 2018.
The results of the EIS improved calibration will be published in 
Del Zanna and Warren (2022, in preparation).

{

\begin{figure}
\centering
\includegraphics[angle=0,width=0.49\hsize,viewport=20 0 498 498,clip]{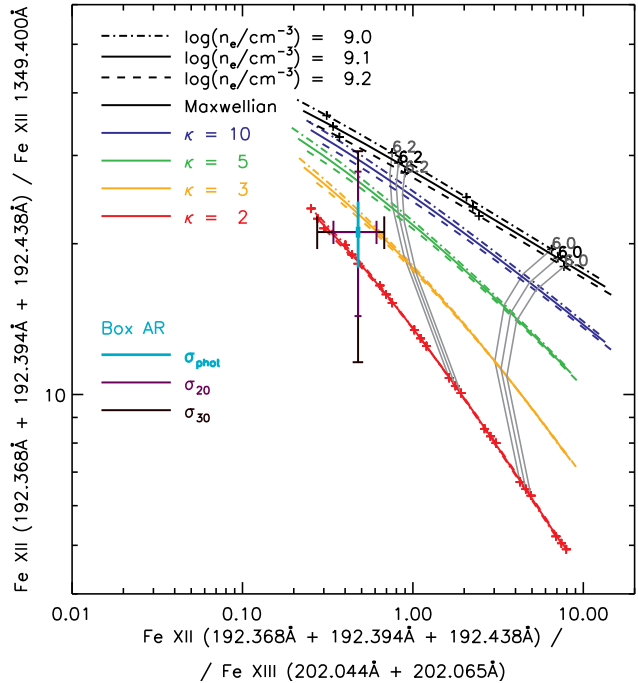}
\caption{Dependence of the NMED diagnostic diagrams on log($N_\mathrm{e}$\,[cm$^{-3}$]). The diagnostic curves shown by full lines and the observed ratios correspond to Figure \ref{fig:nmed} right. Dashed a dot-dashed lines denote changes in electron density equal to 0.1 dex.}
\label{fig:nmed_dens}
\end{figure}
\section{Electron density uncertainties and the diagnostics of $\kappa$}
\label{Appendix:dens_tk}
The measurements of $\kappa$ done in Section \ref{sec:nmed_measurements} required prior determination of electron density. However, the electron densities is also subject to uncertainties of the measured line intensities, especially the photon noise. Note we do not consider the calibration uncertainty, since all lines used for measurements of $N_\mathrm{e}$ are observed by the same channel of EIS.

The photon noise uncertainties are shown by gray stripes in the emissivity ratio plots on Figure \ref{fig:fe_12_em}. It is seen that the photon noise introduces uncertainty into the measurements of log($N_
\mathrm{e}$\,[cm$^{-3}$]) of about 0.1 dex for \ion{Fe}{13}, and slightly larger, $\approx$0.15 dex, for the \ion{Fe}{12} 186.9\,\AA~and 192.4\,\AA~pair of lines.

In Figure \ref{fig:nmed_dens}, we show the changes in the diagnostic diagram for the box AR (see also right panel Figure \ref{fig:nmed}) that occur due to the 0.1 dex uncertainty in the measurements of log($N_\mathrm{e}$\,[cm$^{-3}$]). This uncertainty in electron density is shown by different linestyles. It is seen that for the smallest considered value of $\kappa$\,=\,2, the difference is negligible, while uncertainty in $\kappa$ slightly increases with increasing $\kappa$ (i.e., approaching Maxwellian). However, even then, the curves for $\kappa$\,=\,10 and Maxwellian still do not overlap. Therefore, our determination that the NMED represent a viable explanation of the \ion{Fe}{12} 192.4\,\AA\,/\,1349\,\AA~ratio observed in the AR is not influenced in the uncertainties in the measurements of electron density.
}

\end{document}